\documentclass[]{aastex631}

\usepackage{txfonts}
\usepackage{graphicx}
\usepackage{color}
\usepackage{appendix}

\newcommand{\logt}{log$(T$\,[K]$)$}
\newcommand{\logne}{log$(N_\mathrm{e}$\,[cm$^{-3}$]$)$}
\newcommand{\dens}{$N_\mathrm{e}$}

\definecolor{dg}{rgb}{0.0, 0.5, 0.0}
\definecolor{applegreen}{rgb}{0.55, 0.71, 0.0}
\definecolor{dr}{rgb}{0.75, 0.1, 0.0}
\definecolor{orange}{rgb}{0.9, 0.5, 0.0}
\definecolor{db}{rgb}{0.0, 0.3, 0.8}

\begin{document}

\title{KAPPA: A Package for Synthesis of Optically Thin Spectra for the Non-Maxwellian $\kappa$-Distributions. \\ III. Improvements to Ionization Equilibrium and Extension to $\kappa$\,$<$\,2}

\correspondingauthor{Jaroslav Dud\'{i}k}
\email{jaroslav.dudik@asu.cas.cz}
\affil{Astronomical Institute of the Czech Academy of Sciences, Fri\v{c}ova 298, 251 65 Ond\v{r}ejov, Czech Republic}

\author[0000-0003-2629-6201]{Elena Dzif\v{c}\'{a}kov\'{a}}
\affil{Astronomical Institute of the Czech Academy of Sciences, Fri\v{c}ova 298, 251 65 Ond\v{r}ejov, Czech Republic}

\author[0000-0003-1308-7427]{Jaroslav Dud\'{i}k}
\affil{Astronomical Institute of the Czech Academy of Sciences, Fri\v{c}ova 298, 251 65 Ond\v{r}ejov, Czech Republic}

\author[0009-0007-7715-9676]{Martina Pavelkov\'{a}}
\affil{Astronomical Institute of the Czech Academy of Sciences, Fri\v{c}ova 298, 251 65 Ond\v{r}ejov, Czech Republic}

\author[0009-0005-1782-6671]{Bo\v{z}ena Solarov\'{a}}
\affil{Astronomical Institute of the Czech Academy of Sciences, Fri\v{c}ova 298, 251 65 Ond\v{r}ejov, Czech Republic}

\author[0000-0002-7565-5437]{Alena Zemanov\'{a}}
\affil{Astronomical Institute of the Czech Academy of Sciences, Fri\v{c}ova 298, 251 65 Ond\v{r}ejov, Czech Republic}

\begin{abstract}
The KAPPA package is designed for calculations of optically thin spectra for the non-Maxwellian $\kappa$-distributions. This paper presents extension of the database to allow calculations of the spectra for extreme values of $\kappa$\,$<$\,2, which are important for accurate diagnostics of the $\kappa$-distributions in the outer solar atmosphere. In addition, two improvements were made to the ionization equilibrium calculations within the database. First, the ionization equilibrium calculations now include the effect of electron impact multi-ionization (EIMI). Although relatively unimportant for Maxwellian distribution, the EIMI becomes important for some elements such as Fe and low values of $\kappa$, where it modifies the ionization equilibrium significantly. Second, the KAPPA database now includes the suppression of dielectronic recombination at high electron densities, evaluated via the suppression factors. We find that at the same temperature, the suppression of dielectronic recombination is almost independent of $\kappa$. The ionization equilibrium calculations for the $\kappa$-distributions are now provided for a range of electron densities.
\end{abstract}


%
%_________________________________________
\section{Introduction}
\label{Sect:Intro}

The non-Maxwellian $\kappa$-distributions, characterized by a near-Maxwellian core and a high-energy power-law tail \citep{Oka13,Livadiotis17,Lazar21}, have been detected in a range of astrophysical plasmas. First detected in planetary magnetospheres  \citep[e.g.,][]{Vasyliunas68a,Vasyliunas68b,Olbert68,Dialynas09,Dialynas18,Kirpichev21}, the $\kappa$-distributions also provide a good fit to the halo and strahl components of the solar wind \citep[the literature is extensive, but see, e.g.,][]{Maksimovic97a,Maksimovic97b,Stverak08,LeChat09,Pierrard16,Bercic19,Wilson19,Scherer22a,Scherer22b}, or the plasma of the interplanetary CMEs \citep{Gu23}. The $\kappa$-distributions were also detected in the cometary coma \citep{Broiles16}. Recent reviews on the presence and consequences of $\kappa$-distributions in the solar wind, heliosphere, and planetary magnetospheres are provided for example by \citet{Pierrard21} and \citet{Maksimovic21}. The $\kappa$-distributions were also proposed to explain observed emission spectra of planetary nebulae \citep[e.g.,][]{Nicholls12,Nicholls13,Yao22}, and subsequently also galactic jets \citep{Davelaar19} and active galactic nuclei \citep{Morais21}; although detection of weak departures from Maxwellian in the planetary nebulae requires reliable atomic data \citep[see][]{Storey13,Storey15,Storey14}. Finally, other indications of non-Maxwellian particles with high energies were also obtained in supernova remnants \citep[e.g.,][]{Raymond10,Raymond17}.

Detection of the $\kappa$-distributions in the low solar corona or transition-region (including source regions of the solar wind) relies on analyses of emission spectra. In principle, electron $\kappa$-distributions can be detected from ratios of emission line intensities, while ion $\kappa$-distributions can be detected by analysis of well-resolved emission line profiles. Indications of the presence of electron $\kappa$-distributions from line intensity ratios were obtained for example by \citet{Dudik15}, \citet{Lorincik20}, and \citet{DelZanna22}, who showed that in active regions, the distribution is likely strongly non-Maxwellian with a very low value of $\kappa$\,$\approx$\,2. Similarly, emission lines from flaring plasma also show strong departures from Maxwellian \citep{Dzifcakova18}. Contrary to that, the quiet Sun or even bright point corona are Maxwellian \citep{Lorincik20,DelZanna22,Savage23}. In \citet{DelZanna22}, the quiet Sun was observed within the same dataset directly in the vicinity of the active region. In \citet{Lorincik20}, the quiet Sun was observed at a date similar as the active region. Therefore, in both cases the degradation of the instrument sensitivity with time could have been neglected. This is important, since the spatial variations of $\kappa$ are therefore independent of instrument calibration used. Thus, the veracity of detection of $\kappa$-distributions in active region corona can be established. Emission line profiles from the transition-region, corona, and flaring plasma are also consistent with being strongly non-Maxwellian with low values of $\kappa$ \citep{Jeffrey16,Jeffrey17,Jeffrey18,Dudik17,Polito18}. In addition, some continuum bremsstrahlung emission from flaring plasma was also found to be consistent with $\kappa$-distributions in some instances \citep[][]{Kasparova09,Oka13,Oka15,Battaglia15,Battaglia19}. Finally, other indications of the presence of $\kappa$-distributions in solar flares were obtained from hydrodynamic modelling of various spectral properties \citep[see][]{Allred22}.

Recently, \citet{Mondal20} reported presence of weak radio bursts (in the mSFU and sub-picoflare range) originating in the solar corona. Such weak bursts are likely associated with small EUV brightenings \citep{Mondal21}. Subsequently, \citet{Sharma22} showed that such radio events are ubiquitous, occur both in the quiet Sun and active regions, carry sufficient energy to heat the solar corona, and are associated with acceleration of electrons to 0.4--4\,keV. The more energetic events were found in the vicinity of active regions. We note that these electron energies are similar to the electron energies possessed by the high-energy tail of the $\kappa$-distributions detected from analysis of emission line spectra \citep[see, e.g., the top panel of Figure 10 in][]{DelZanna22}. We also note that the energies of the bursts detected by \citet{Mondal20} are possibly too low to be detected with dedicated hard X-ray instrumentation used to observe energies only above 2--4\,keV depending on the instrument \citep[see, e.g.,][]{Hannah10,Hannah16,Marsh17,Buitrago-Casas22,Paterson23}, while focusing on constraining the non-thermal electrons in the quiet Sun.

The fundamental reason for the presence of high-energy tails in astrophysical plasmas is the $E^{-2}$ behavior of Coulomb collision cross-section with the particle kinetic energy $E$, meaning that progressively high-energy particles become less collisional \citep{Scudder79}. Consequently, the solar corona above about 1.05\,$R_\odot$ is expected to be non-Maxwellian \citep{Scudder13,Scudder19}. Particle acceleration resulting in high-energy tails can also occur via multiple processes, including turbulence \citep[where the inverse proportionality of turbulent diffusion coefficient to velocity leads directly to $\kappa$-distributions, see][]{Hasegawa85,Laming07,Bian14,Demaerel20}, magnetic reconnection \citep[e.g.,][]{Gontikakis13,Ripperda17,Threlfall18,Arnold21,Oka22}, wave-particle interactions \citep{Vocks08,Vocks16}, or steep gradients of temperature and density, such as in the transition region \citep[e.g.,][]{Roussel-Dupre80a,Ljepojevic88,Dzifcakova17}. Furthermore, physical correlations between particles specifically lead to $\kappa$-distributions \citep[see][and references therein]{Livadiotis22}.

The spectroscopic diagnostics of $\kappa$-distributions in the outer solar atmosphere rely on the availability of spectral synthesis, which in turn relies on availability of the atomic datasets containing rates for individual processes for $\kappa$-distributions. This task is accomplished by the KAPPA database and software\footnote{\url{kappa.asu.cas.cz}}. In Paper I, \citep{Dzifcakova15}, the basic concept was established and presented. In Paper II \citep{Dzifcakova21}, the database was updated to be compatible with the latest release of CHIANTI version 10 \citep{Dere97,DelZanna21}, including additional processes such as two-ion model. Here, we provide improvements to the existing database and software by including calculations for values of $\kappa$\,$<$\,2 (Sections \ref{Sect:Ioniz_low_kappa} and \ref{Sect:Excit_rates}), as well as processes such as electron impact multi-ionization (Section \ref{Sect:Ioniz_EIMI}) and density suppression of dielectronic recombination (Section \ref{Sect:Ioniz_DR_suppression}). Finally, we note that since 2020 the synthetic spectra for non-Maxwellian $\kappa$-distributions can also be calculated using the AtomDB project and its extensions \citep[see][]{Smith01,Foster12,Cui19,Foster20}. The AtomDB project relies on the Maxwellian decomposition method \citep[][see also our Section \ref{Sect:Ioniz_DR_suppression}]{Hahn15a}, while the KAPPA package relies on its own calculations of the individual non-Maxwellian rates (see Paper I and II for details) and follows the CHIANTI database formats and procedures.

%
%
%_________________________________________
\section{The Non-Maxwellian Electron $\kappa$-Distributions}
\label{Sect:Kappa}

\begin{figure}[t]
  \centering     
    \includegraphics[width=8.8cm]{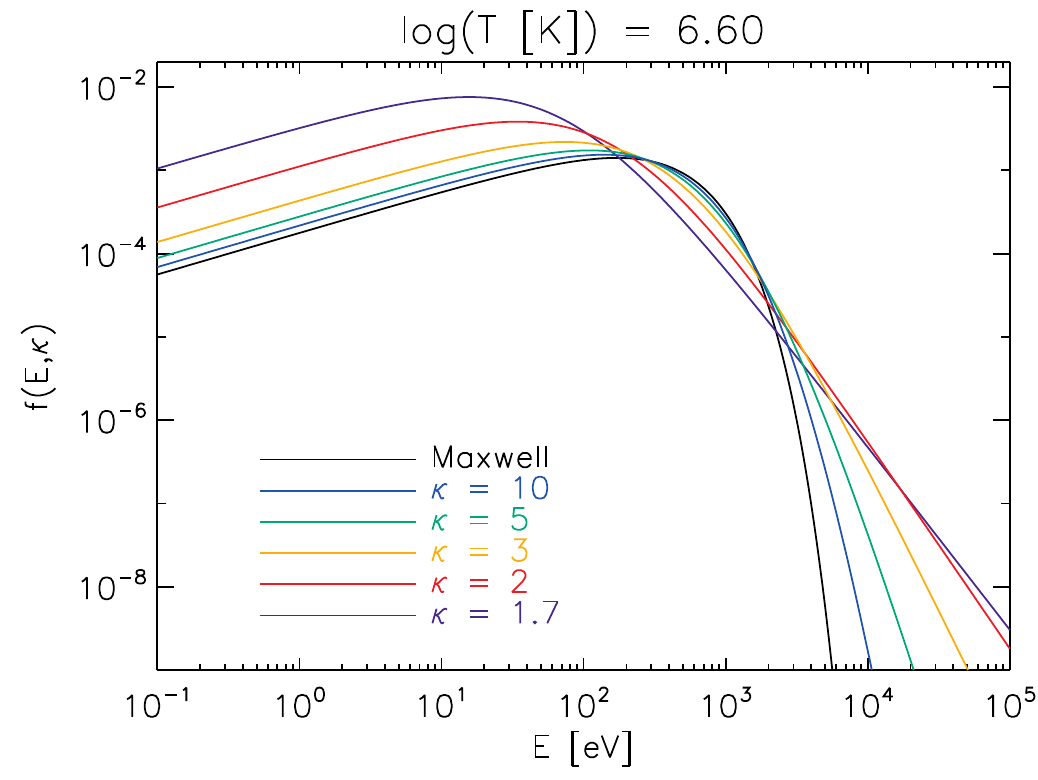}
  \caption{Electron $\kappa$-distributions as a function of electron kinetic energy $E$, plotted for various values of $\kappa$ and \logt\,=\,6.6.}
  \label{Fig:Kappa_distribution}
\end{figure}

The KAPPA database allows for calculations of the synthetic optically thin spectra for the non-Maxwellian electron $\kappa$-distributions \citep{Olbert68,Vasyliunas68a,Vasyliunas68b,Livadiotis09}. We note that several definitions of the $\kappa$-distributions exist \citep[see, e.g.,][]{Livadiotis13,Livadiotis17,Lazar16,Lazar21}. At present, the KAPPA database uses a simple formulation for the isotropic electron distribution $f_\kappa(E)dE$, which depends only on the electron kinetic energy $E$, and has two parameters, $\kappa$ and $T$:
\begin{equation}
	f_\kappa(E)dE= A_\kappa  \frac{2}{\sqrt{\pi}\left(k_\mathrm{B}T\right)^{3/2}} \frac{E^{1/2}dE}{\left(1+\frac{E}{(\kappa-3/2)k_\mathrm{B}T}\right)^{\kappa+1}}\,,
 \label{Eq:Kappa}
\end{equation}
where $k_\mathrm{B}$ is the Boltzmann constant, $A_\kappa$ is a normalization constant, and the mean kinetic energy $\left<E\right>$ is given by the expression $\left<E\right> = 3/2 k_\mathrm{B}T$. The $3/2 < \kappa < \infty$ is an independent parameter describing the degree of departure from Maxwellian, which corresponds to the limit of $\kappa$\,$\to$\,$\infty$ (see Figure \ref{Fig:Kappa_distribution}). We also note that the Equation \ref{Eq:Kappa} corresponds to a "Kappa--A" distribution as defined and discussed by \citet{Lazar16}.

At present, such definition of $\kappa$-distribution is sufficient to evaluate the effects of the non-Maxwellian electron distributions (NMED) characterized by a high-energy power-law tail on the optically thin spectra. This is because for the $\kappa$-distributions, the tail is relatively strong (see Figure \ref{Fig:Kappa_distribution}) compared for example to the distributions composed of a core Maxwellian and a power-law tail \citep{Dzifcakova11b}, used to describe the electron distribution derived from X-ray bremsstrahlung in flares. Still, even with a relatively strong tail, the changes in intensity of most emission lines with $\kappa$-distributions are of the order of several tens of per cent, and rarely more than a factor of two compared to the Maxwellian \citep[see, e.g.,][]{Dzifcakova10,Mackovjak13,Dudik15,Lorincik20,DelZanna22,Savage23}. The reasons for this choice are chiefly twofold. First, the observational uncertainties in the emission line intensities are relatively large due to the radiometric calibration and its degradation, with the precision of the calibration of spectroscopic instruments such as Hinode/EIS being about 20\% \citep[see, e.g.,][]{Culhane07,BenMoussa13,DelZanna13calib}. Second, the $\kappa$-distributions (and any other NMED) are usually detectable using line intensity ratios involving two lines with different sensitivity to $\kappa$, such as an allowed and a forbidden line \citep[see, e.g.,][]{Dudik17Rev,Dzifcakova18,Lorincik20,DelZanna22}. Forbidden lines are usually much weaker than allowed ones, which means their photon noise uncertainty also limits the determination of $\kappa$ only to a range of values. Therefore, resolving different types of $\kappa$-distributions or indeed differentiating between $\kappa$-distributions and other NMED with a high-energy power-law tail in the outer solar atmosphere is very difficult at present.

Finally, we note that we do not repeat here the equations for calculation of the line intensities or individual ionization, recombination, and excitation coefficients, as these have been presented and discussed at length in previous literature, and are summarized in the previous Papers I and II. In the following, we describe the upgrades to the existing database and software.

%
%
%%-------------------------- FIGURE 1
%%--------------------------
%%
%
%
%%-------------------------- FIGURE 2
\begin{figure*}[t]
  \centering     
  \includegraphics[width=8.8cm, clip, viewport=20 10 410 300]{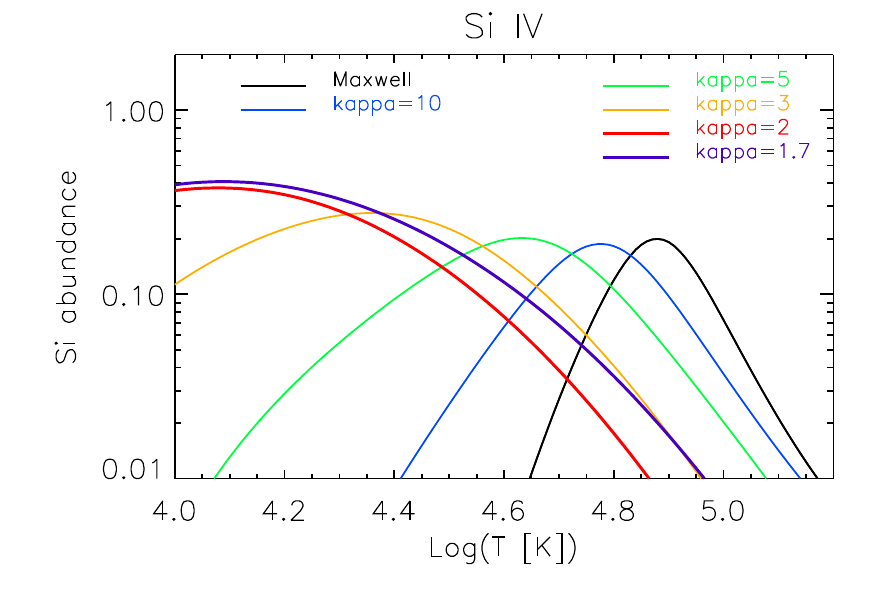}
  \includegraphics[width=8.8cm, clip, viewport=20 10 410 300]{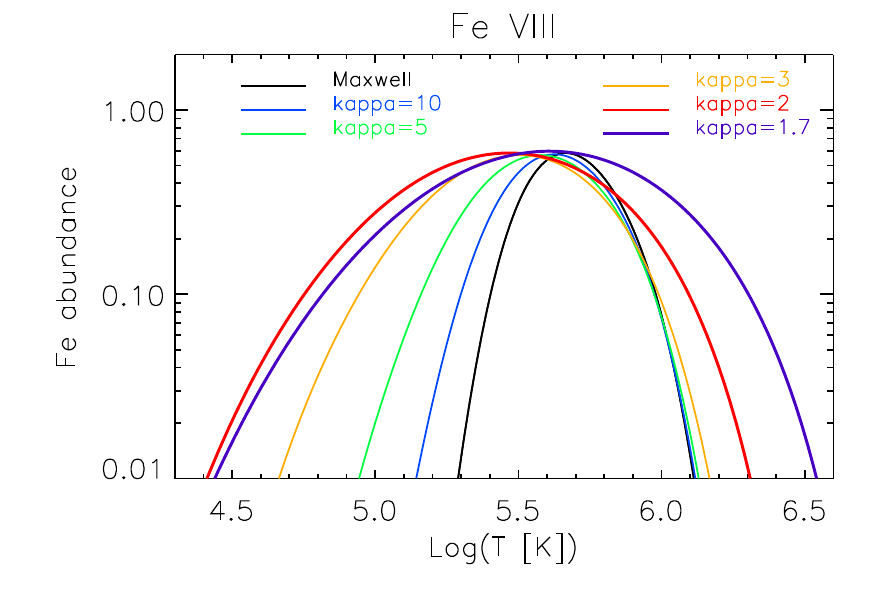}
  \includegraphics[width=8.8cm, clip, viewport=20 10 410 300]{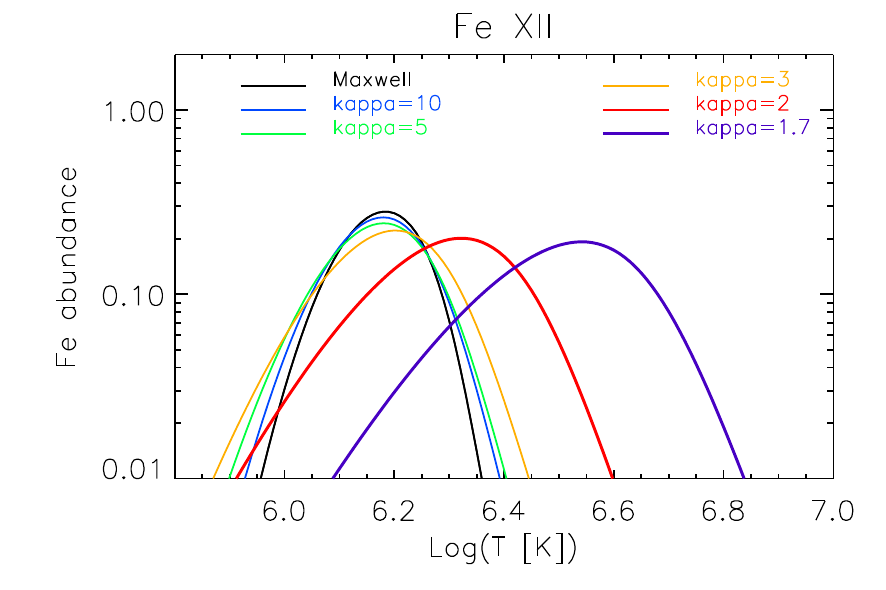}
  \includegraphics[width=8.8cm, clip, viewport=20 10 410 300]{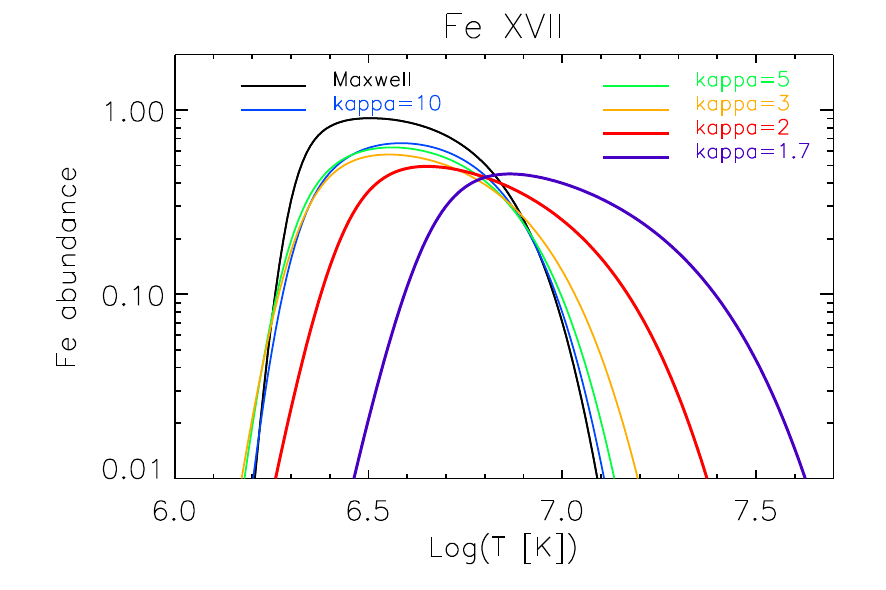}
  \caption{Relative ion abundances and their changes with decreasing $\kappa$. Individual colors denote the value of $\kappa$. Violet stands for $\kappa$\,=\,1.7, while red denotes $\kappa$\,=\,2. A variety of ions is shown, from the transition region \ion{Si}{4} and \ion{Fe}{8} to the coronal \ion{Fe}{12} and \ion{Fe}{17}. 
  }
  \label{Fig:Fe_low_k}
\end{figure*}
%%--------------------------
%
%
%%-------------------------- FIGURE 3               ONE COLUMN FIGURE, should have width=8.8cm. 
%                                                   width=9.0cm is used to force one-column figure in "linenumbers" style
\begin{figure}[t]
  \centering     
    \includegraphics[width=9.0cm, clip, viewport=12 10 555 340]{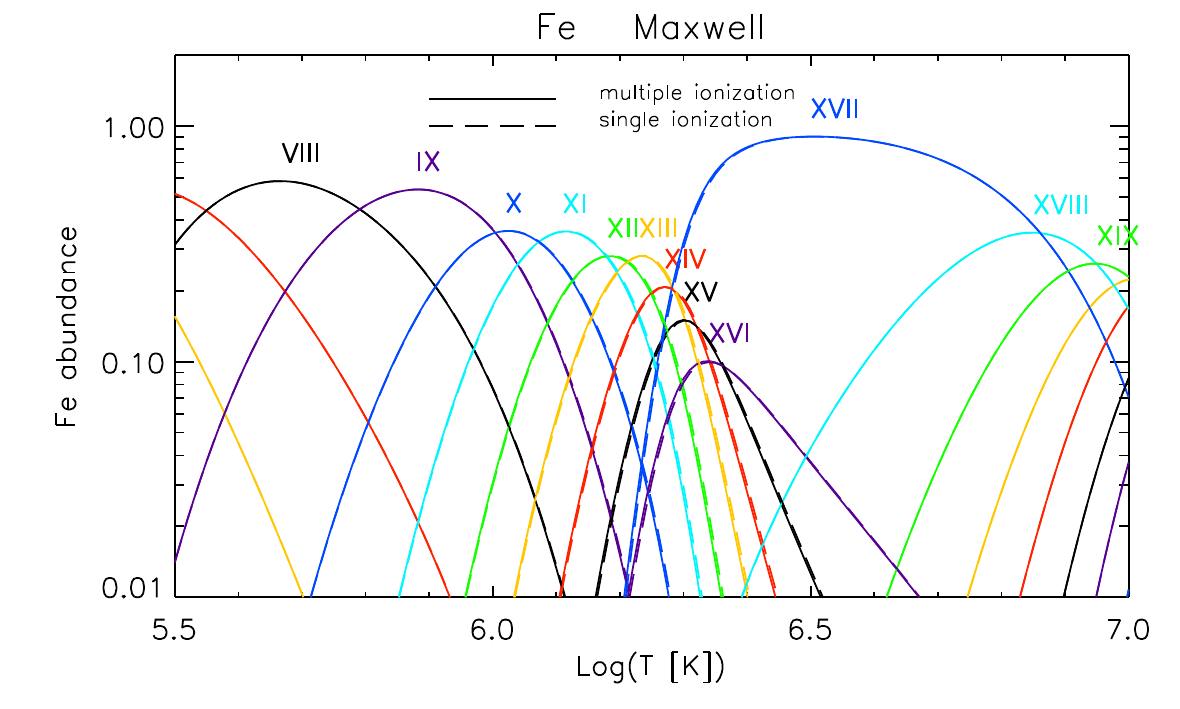}
    \includegraphics[width=9.0cm, clip, viewport=12 10 555 340]{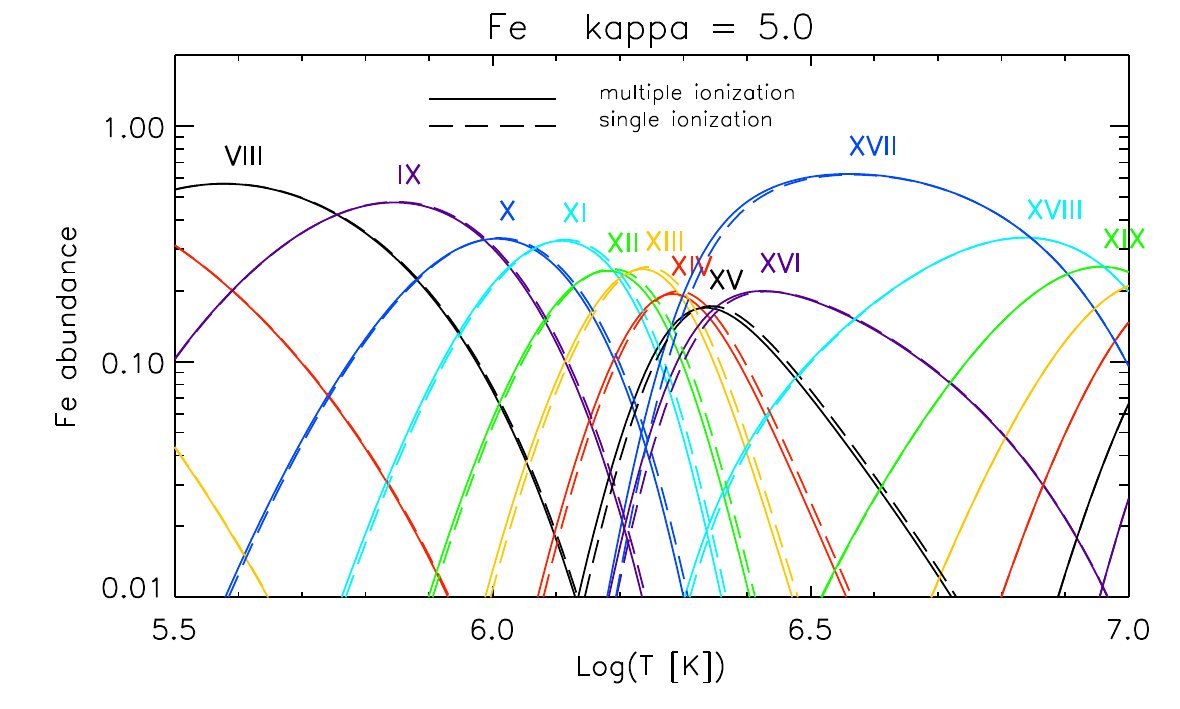}
    \includegraphics[width=9.0cm, clip, viewport=12 10 555 340]{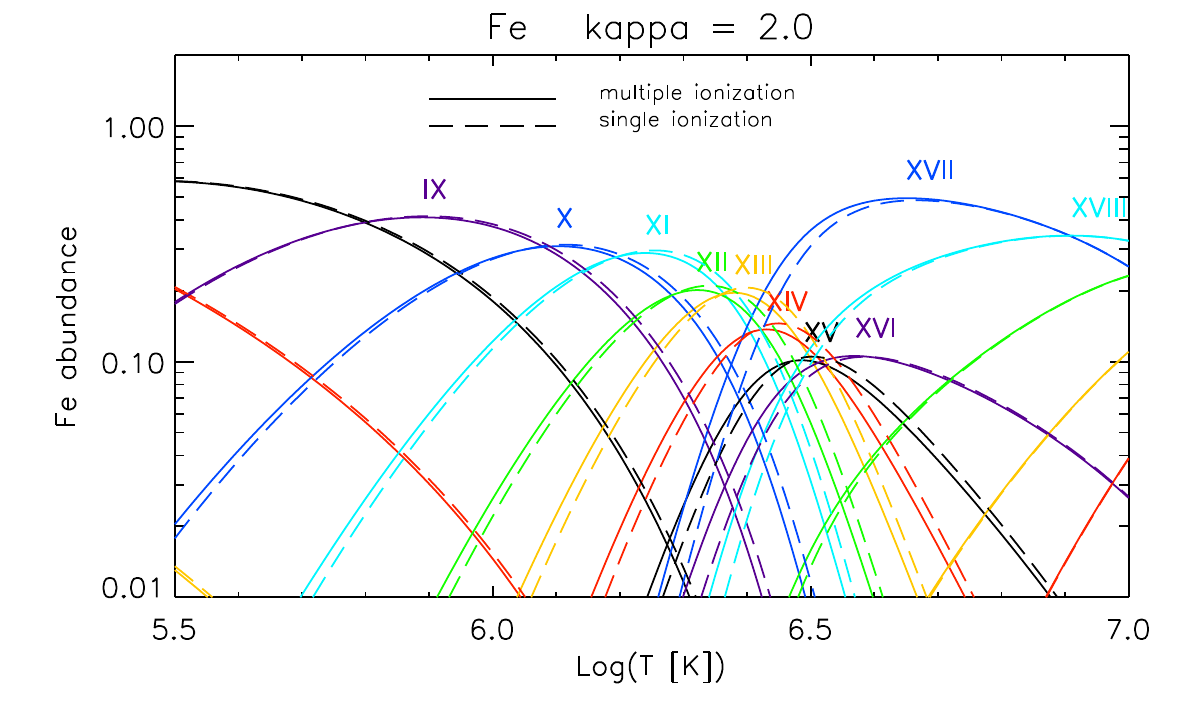}
    \includegraphics[width=9.0cm, clip, viewport=12 10 555 340]{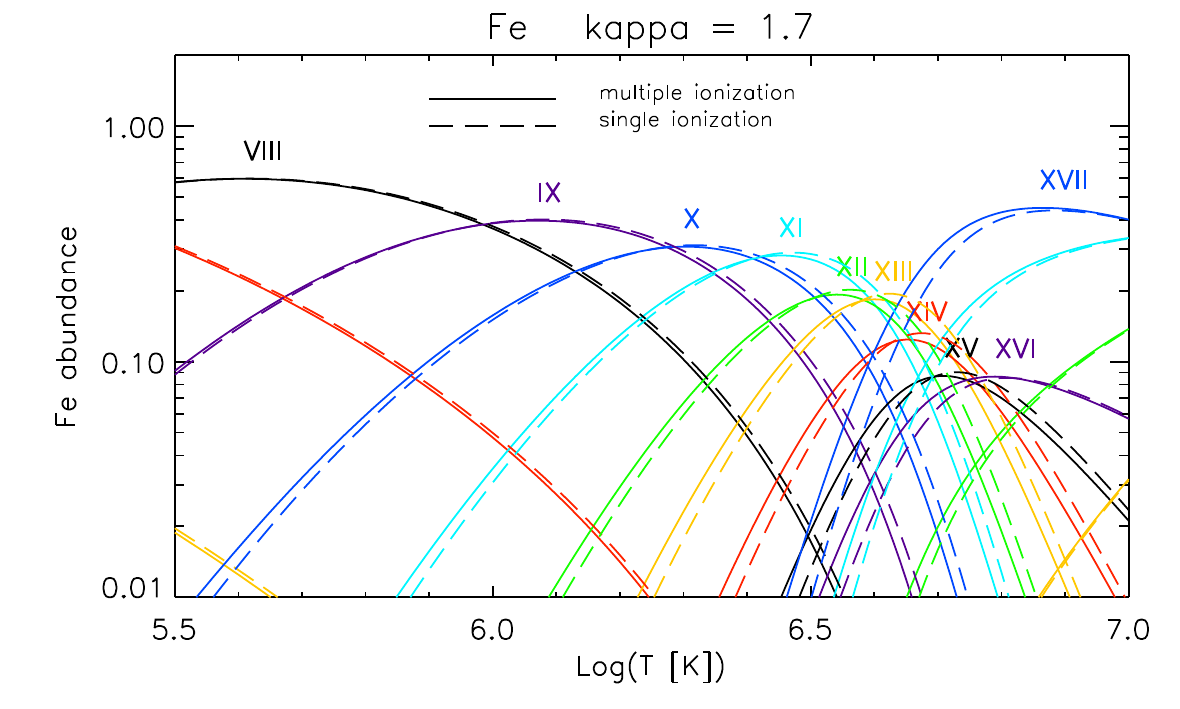}
  \caption{The effect of multi-ionization on the Fe ionization equilibrium for Maxwellian distribution (top) and $\kappa$-distributions with $\kappa$=5, 2, and $\kappa$=1.7 (rows 2--4). Calculations for all distributions, including the Maxwellian, were based on the ionization cross-sections of \citet{Hahn17} and recombination rates from CHIANTI version 10.1.}
  \label{Fig:fe_singl_multi}
\end{figure}
%%--------------------------
%

%
%
%
%
%_________________________________________
\section{Improvements in Ionization and Recombination Rate Calculations}
\label{Sect:Ioniz}

%
%-------------------------------
\subsection{Ionization and Recombination Rates for $\kappa < 2$}
\label{Sect:Ioniz_low_kappa}

Previous versions of the KAPPA database contained invididual rates for discrete values of $\kappa$\,=\,2, 3, 4, 5, 7, 10, 15, 25, and 33 (see Paper I). However, recent diagnostics of electron distribution in the solar corona showed that the value of $\kappa$ can be lower than two \citep[see][]{Dudik15,Dudik17,Dzifcakova18,Polito18,Lorincik20,DelZanna22}. Previously, the value of $\kappa$\,=\,2 was considered a relatively extreme one and was the lowest one available in the KAPPA database. We now extended our calculations of individual rates, as well as ionization equilibria for values of $\kappa < 2$. Namely, the newly available values are $\kappa$\,=\,1.9, 1.8, and 1.7. We note that the asymptotic value of $\kappa$ is $\kappa$\,$\to$\,1.5. However, our choice of $\kappa$\,=\,1.7 are about the lowest detected \citep{Dudik17,Polito18} and should be sufficient to illustrate the effects of such low $\kappa$-distributions on the spectra, taking into account the increasing uncertainty of the approximations used for the calculation of recombination and excitation rates \citep{Dzifcakova15}. The value of $\kappa$\,=\,2.5 was also added to bridge the relatively large gap between $\kappa$\,=\,2 and 3. This value of $\kappa$ also corresponds approximately to one of the critical $\kappa$ indices in non-extensive thermodynamics \citep[see][]{LIvadiotis10}.

We calculated ionization rates directly from the cross-sections, similarly as for other values of $\kappa$ (see Paper II). The cross-sections we used are those from the compilation of \citet{Hahn17}, similarly as the latest version 10.1 of CHIANTI \citep[][]{Dere23}. The recombination rates were calculated using the approximate method of \citet{Dzifcakova92} and \citet{Dzifcakova13}, and subsequently obtained the ionization equilibria. We note that the ionization equilibria for such low $\kappa$ values were for the first time calculated by \citet{Hahn15a}, and we checked that our calculations are in good agreement with those using the method of \citet{Hahn15a}. 

The changes in the ionization equilibrium of iron for low $\kappa$ in comparison with $\kappa$\,=\,2 and other values of $\kappa$ are shown in Figure \ref{Fig:Fe_low_k}. For transition region ions such as \ion{Si}{4} and \ion{Fe}{8} (top row); that is, for temperatures below about \logt\,$\approx$\,5.7, the changes in the peaks of the relative ion abundance for $\kappa$\,=\,1.7 are small compared to $\kappa$\,=\,2. The peaks are widened and shifted to slightly higher temperatures. Note that this shift is in reverse direction than shifts for higher $\kappa$ values \citep[cf.,][]{Dzifcakova13}. That is, with decreasing $\kappa$, the peaks are first shifted to progressively lower $T$ until about $\kappa$\,=\,2; then, for $\kappa$\,$<$\,2, the peaks shift to slightly higher $T$. This effect is well visible for \ion{Fe}{8} (see the top right panel of Figure \ref{Fig:Fe_low_k}).

At coronal temperatures and higher degrees of ionization, a small change of $\kappa$ means substantial changes in the ionization peaks (bottom row of Figure \ref{Fig:Fe_low_k}). The shape of peaks is wider, but the relative ion abundances are shifted significantly to higher temperatures. For $\kappa$\,=\,1.7, the peak temperatures can be up to a factor $<$2 higher than for $\kappa$\,=\,2. This happens for both \ion{Fe}{12}, where the shift of the log$(T_\mathrm{max}$\,[K]) is from 6.35 to 6.55 (a factor of $\approx$1.6 higher), as well as for \ion{Fe}{17}, where the ionization peak shifts from log$(T_\mathrm{max}$\,[K])\,=\,6.65 for $\kappa$\,=\,2 to 6.85 for $\kappa$\,=\,1.7 (again a factor of $\approx$1.6). This means that the peak of the relative ion abundance continues its shift to higher temperatures with decreasing $\kappa$ \citep[a fact previously described up to $\kappa$\,=\,2 by][]{Dzifcakova92,Dzifcakova02,Dzifcakova13}. Reliable detection of such extremely low values of $\kappa$ in the solar corona would have significant consequences for the thermal energy content of the solar corona, and thus coronal heating requirements.

%
%-------------------------------
\subsection{Electron Impact Multi-ionization (EIMI)}
\label{Sect:Ioniz_EIMI}

A single electron-ion collision can lead not only to ionization or excitation, but also to multiple ionization of the target ion. For example, it is possible to produce \ion{Fe}{14} directly from \ion{Fe}{12} if the impacting electron has sufficient energy. Triple ionizations are also possible, although the cross-sections fall rapidly with each additional ejected electron \citep[see][]{Hahn17}. Such electron impact multi-ionizations (EIMI) are typically neglected for equilibrium (Maxwellian) plasmas, because the EIMI becomes important only at very high temperatures, where the abundance of the target ion is already small. For iron, EIMI changes the ionization equilibrium (charge state distribution) by less than about 5\%, see Figure 3 of \citet[][]{Hahn15b}. However, EIMI becomes important both in situations when the plasma is rapidly heated (where EIMI reduces the time plasma needs to reach ionization equilibrium), or when the electron distribution is non-Maxwellian with high-energy tails \citep{Hahn15b}. For a $\kappa$-distribution with extremely low $\kappa$, the EIMI changes the relative ion abundances of Fe by a much larger amount, up to a factor of 2--6 depending on the ion and $T$ \citep[see Figure 9 of][]{Hahn15a}. Clearly, the EIMI becomes an important process for such NMED and needs to be taken into account in calculating the ionization equilibrium.

Denoting $I_{km}$ and $R_{mk}$ the ionization and recombination rate coefficients for ions in $k$-th and $m$-th ionization state, $k < m \leq Z$, and $N_k$\,=\,$N(X^{+k})/N(X)$ the relative abundance of the ion $X^{+k}$ and $Z$ is the proton number, the ion populations in equilibrium should satisfy the set of linear equations
\begin{eqnarray}
\nonumber   -(I_{01}+I_{02})N_0+R_{10}N_1+0+...=&0&  \\ 
\nonumber   I_{01}N_0-(R_{10}+I_{12}+I_{13})N_1+R_{21}N_2+0+...=&0& \\ 
            I_{02}N_0+I_{12}N_1-(R_{21}+I_{23}+I_{24})N_2+R_{32}N_3+...=&0& \\ 
\nonumber   ... & & \\
\nonumber   I_{k-2,\,k}N_{k-2}+I_{k-1,\,k}N_{k-1}-(R_{k,k-1}+I_{k,k+1}+I_{k,k+2})N_{k} +R_{k+1,k}N_{k+1}=&0&, \\
\nonumber   ... & & \\
\nonumber   0+...+I_{Z-2,Z}N_{Z-2}+I_{Z-1,Z}N_{Z-1}-R_{Z,Z-1}N_{Z}=&0&,
%\nonumber   0+...+I_{Z-1,Z+1}N_{Z-1}+I_{Z,Z+1}N_{Z}-R_{Z+1,Z}N_{Z+1}=&0&,
%0+0+...+I_{n-2,n}N_{n-2}+I_{n-1,n}N_{n-1}-R_{n,n-1}N_n=&0&, 
\label{Eq:equi}
\end{eqnarray}
Ionization rates for single ionization and for EIMI were calculated using the approximation formulae for ionization cross sections provided by \citet{Hahn17}. The new ionization equilibria in KAPPA database now include the effects of EIMI for all of elements and $\kappa$. 

As pointed out by \citet{Hahn17}, the effect of multiple ionization on the ionization equilibrium for Maxwellian distribution is small, but can become important for the elements with high Z and low $\kappa$-values or high $T$. These effects are shown in Figure \ref{Fig:fe_singl_multi}, where the ionization equilibrium of iron, including EIMI, is plotted for Maxwellian and several values of $\kappa$. These Fe ionization equilibria are compared to calculations without EIMI and in the low-density limit (cf., Section \ref{Sect:Ioniz_DR_suppression}). The effects of EIMI are of importance for coronal Fe ions such as \ion{Fe}{10}--\ion{Fe}{18} between \logt \,$\approx$\,6--7 \citep[cf., Figure 9 of][]{Hahn15a}, although details depend on the ion. Generally, for the affected ions, EIMI shifts the peak formation temperature $T_\mathrm{max}$ towards \textit{lower} values. 
These shifts of $T_\mathrm{max}$ towards lower values of $T$ occur since additional ionization leads to increase of ionization state at a given electron kinetic energy. For some ions, the changes in the relative ion abundances occur at $T > T_\mathrm{max}$; for others at all temperatures (for example, \ion{Fe}{14}, see Figure \ref{Fig:fe_singl_multi}), while for other ions, such as \ion{Fe}{17}, the EIMI affects dominantly temperatures below $T_\mathrm{max}$. Therefore, the EIMI is also a potentially significant process affecting the X-ray spectra such as those observed by MaGIXS \citep{Savage23}, should the X-ray lines of \ion{Fe}{17}--\ion{Fe}{18} be formed in non-Maxwellian conditions.

We note that for coronal ions, the shifts in the ionization equilibrium with $\kappa$ occur in opposite direction than the shifts due to inclusion of EIMI. That is, with decreasing $\kappa$ the peaks are shifted towards higher $T$ compared to Maxwellian calculations. This shift with $\kappa$ is a result of two competing processes. The increase of the ionization rate due to high-energy electrons pushes ionization peaks towards lower $T$, while the increase of the radiative recombination rates due to the excess of electrons at low energies and low $\kappa$ (cf., Figure \ref{Fig:Kappa_distribution}) pushes them to higher temperatures. In addition, changes in dielectronic recombination rates with $\kappa$ also affect the shift. The resulting net effect of $\kappa$-distributions on coronal Fe ions is the shift of their peaks towards higher temperatures, as the changes in the total recombination rate are larger than the changes in the ionization rate at the temperatures where the ion peak occurs \citep[see Figure 2][]{Dzifcakova13}. When EIMI is added, the total ionization rate is increased, resulting in the shift of the ionization peak to lower $T$, i.e., in the opposite direction to the shifts with $\kappa$.

%
%%-------------------------- FIGURE 4 (added in Revision 1)     ONE COLUMN FIGURE, SHOULD HAVE width=8.8cm 
%                                                                 see note for Figure 2
\begin{figure}[t]
  \centering     
    \includegraphics[width=9.0cm, clip, viewport=25 5 555 340]{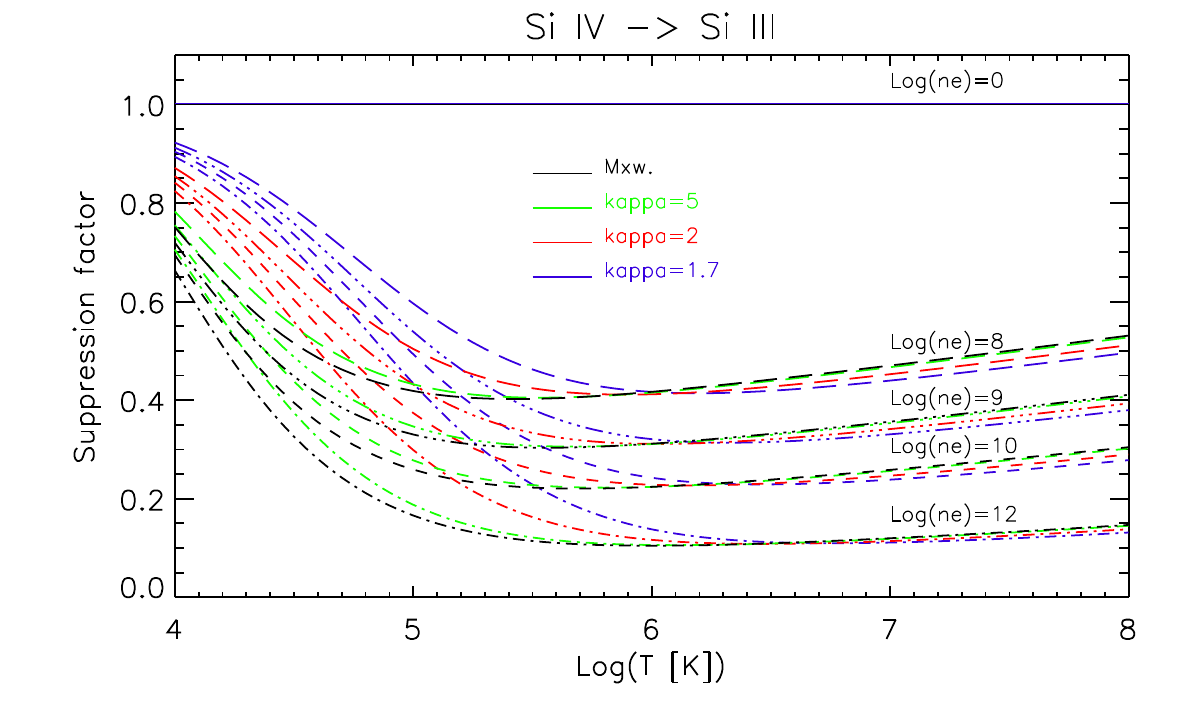}
    \includegraphics[width=9.0cm, clip, viewport=25 5 555 340]{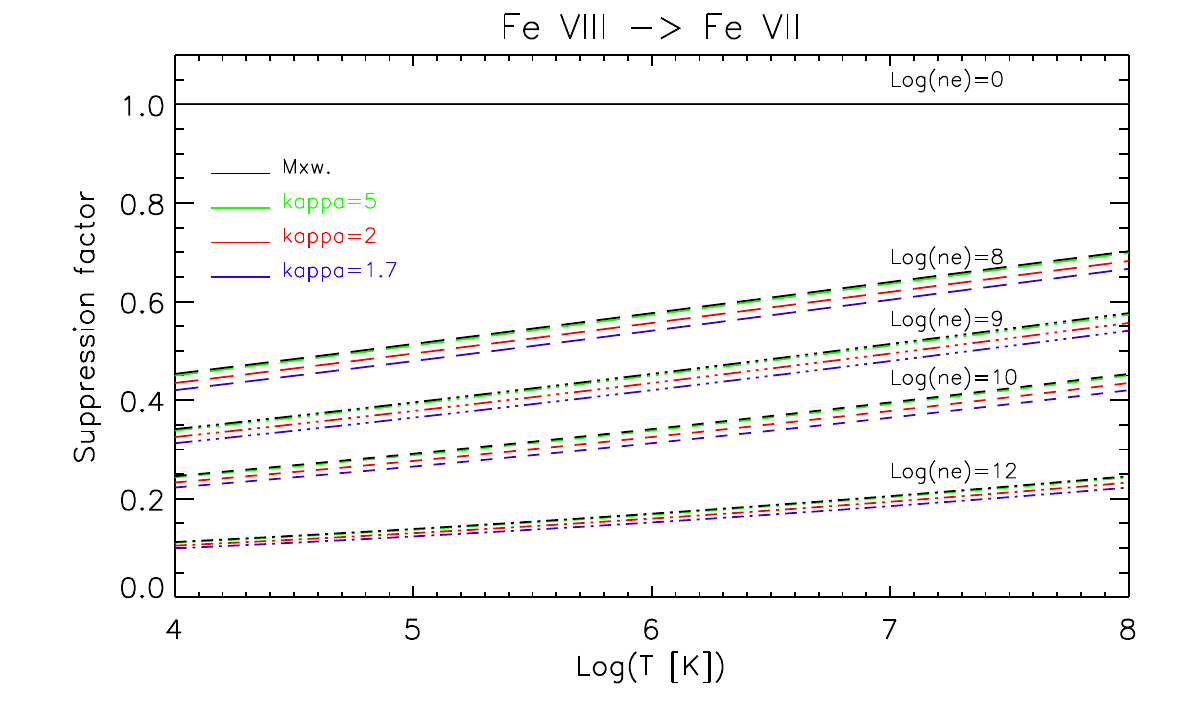}
    \includegraphics[width=9.0cm, clip, viewport=25 5 555 340]{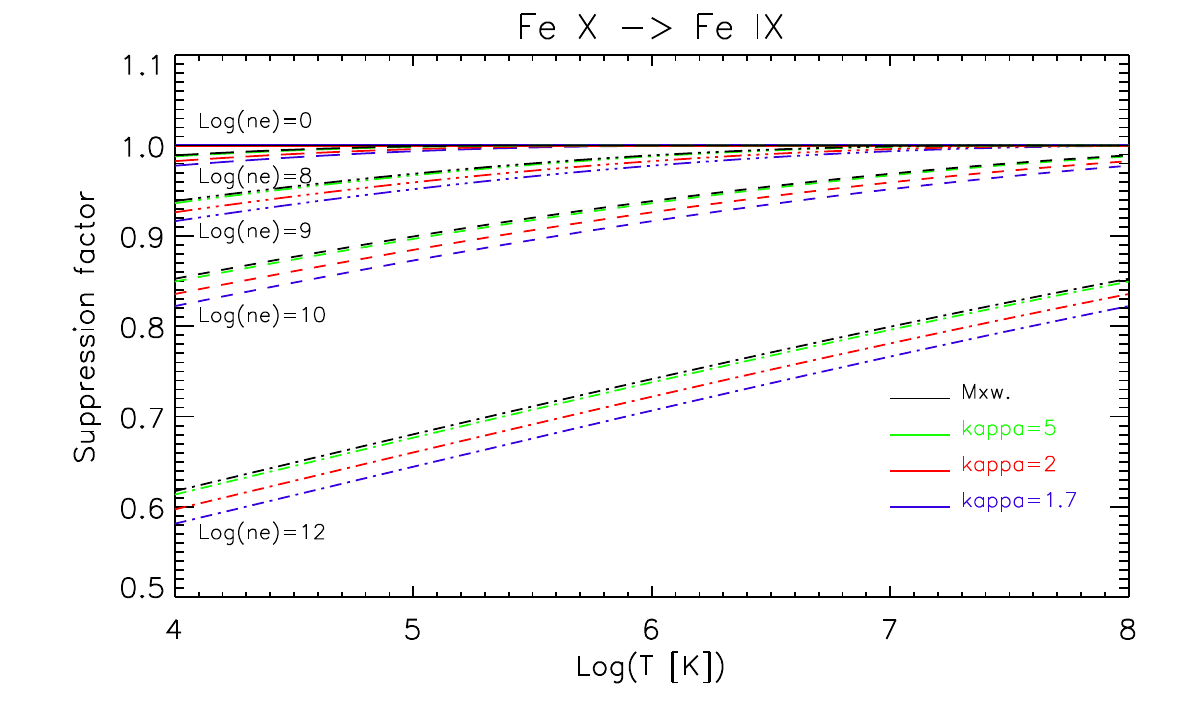}
    \includegraphics[width=9.0cm, clip, viewport=25 5 555 340]{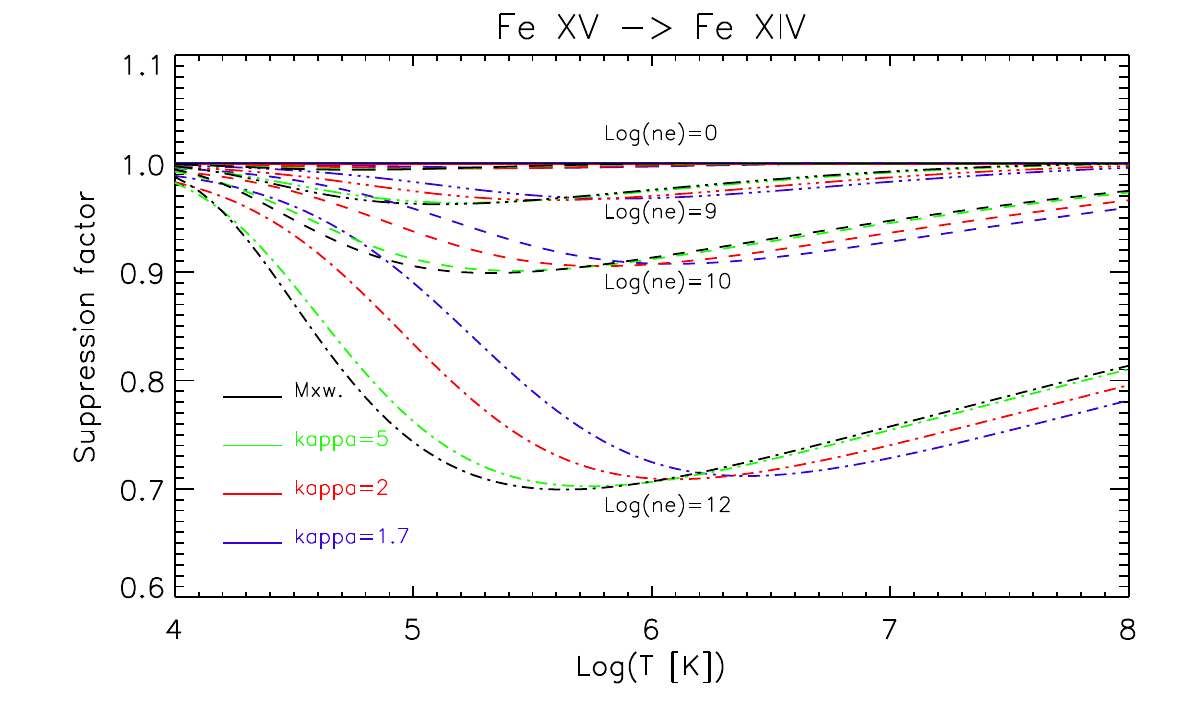}
\caption{Examples of the behavior of the suppression factors for dielectronic recombination for a selection of ions. Individual values of $\kappa$ and \logne~are indicated.}
\label{Fig:dr_sf}
\end{figure}
%%--------------------------
%
%%-------------------------- FIGURE 5 (was Figure 4 in submitted version)
\begin{figure*}[t]
  \centering     
    \includegraphics[width=8.8cm, clip, viewport=5 9 510 340]{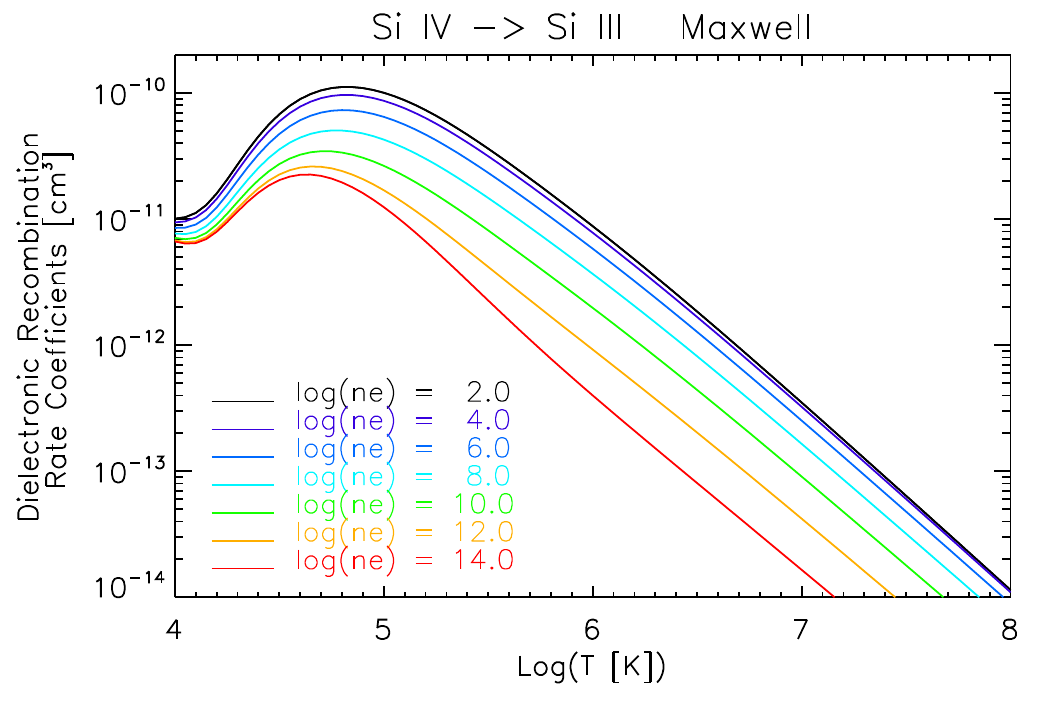}
    \includegraphics[width=8.8cm, clip, viewport=5 9 510 340]{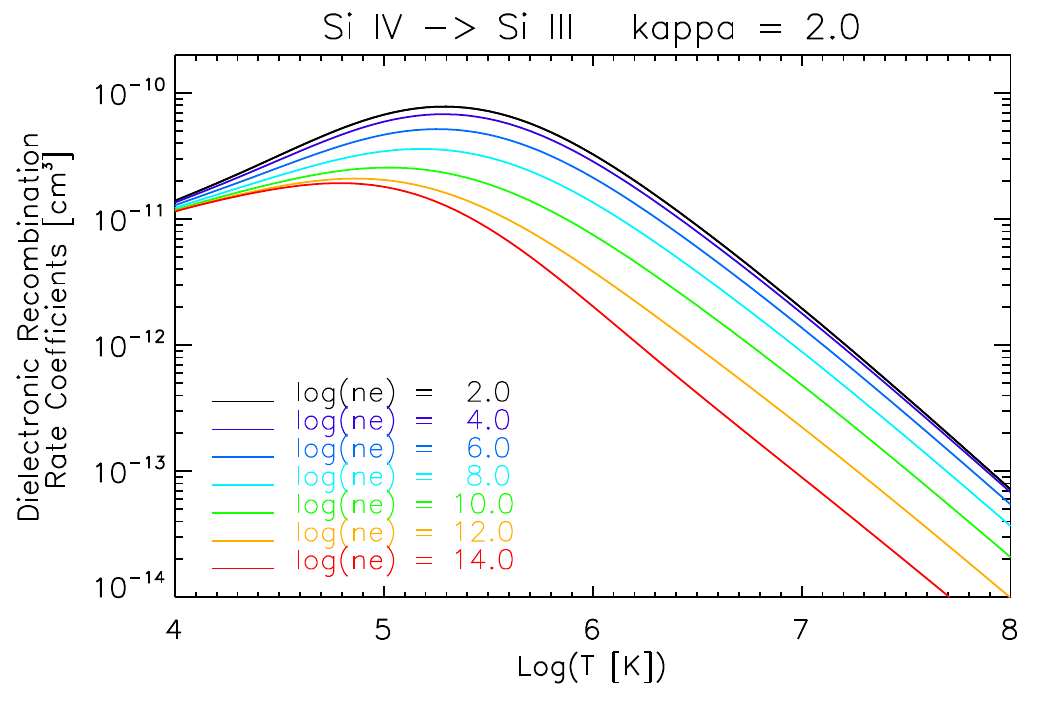}
    \includegraphics[width=8.8cm, clip, viewport=5 9 510 340]{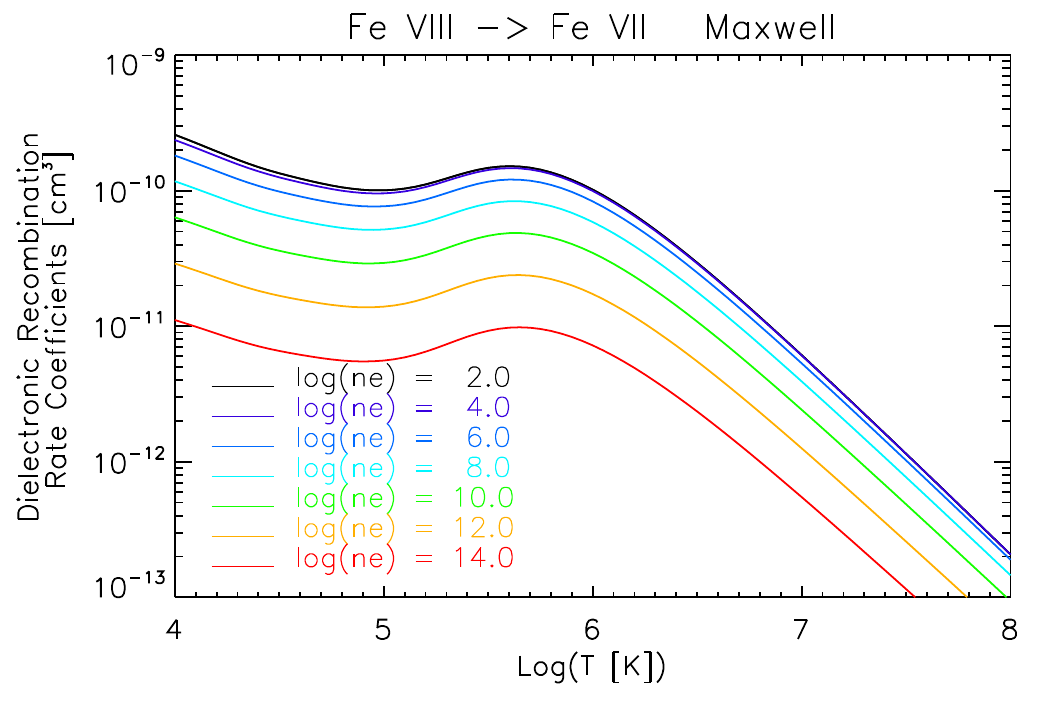}
    \includegraphics[width=8.8cm, clip, viewport=5 9 510 340]{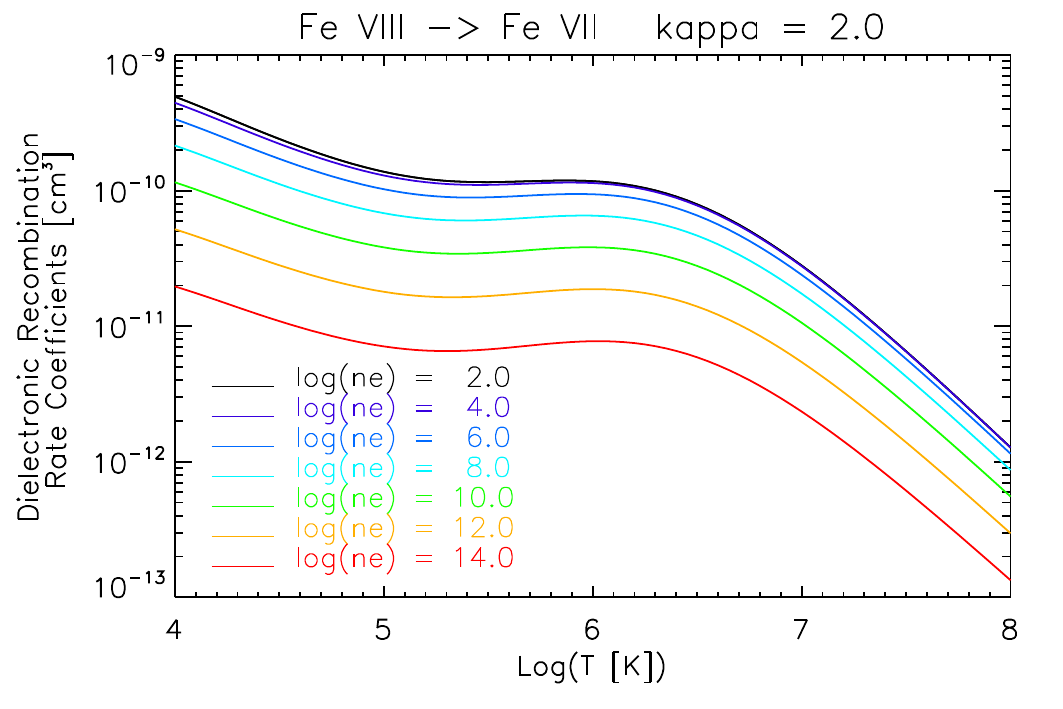}
    \includegraphics[width=8.8cm, clip, viewport=5 9 510 340]{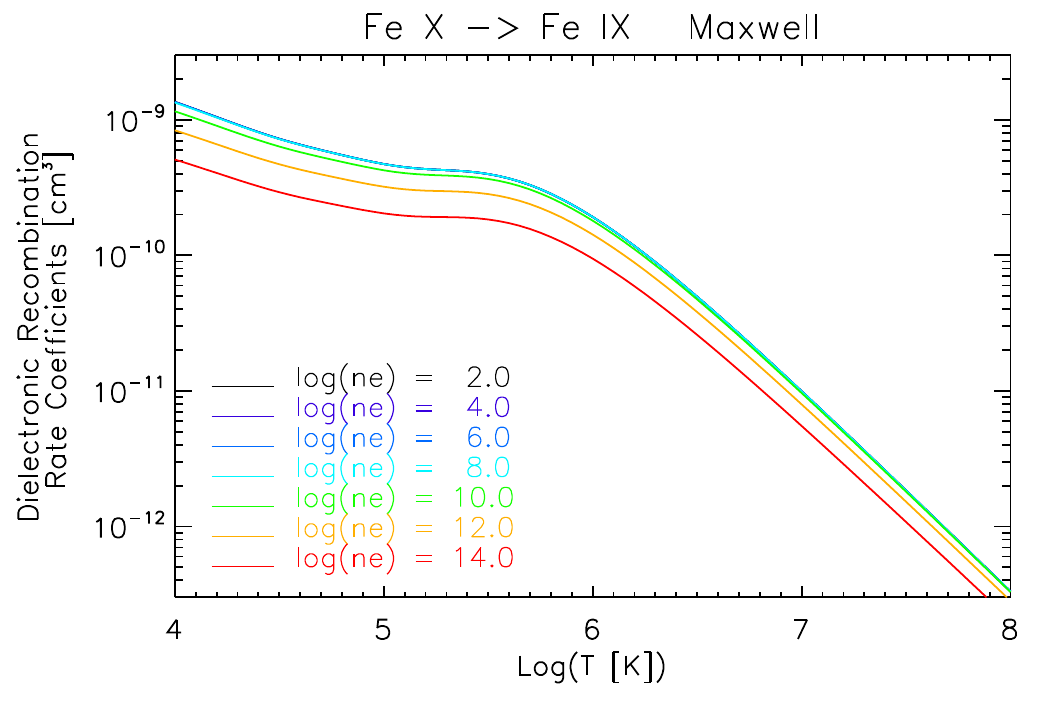}
    \includegraphics[width=8.8cm, clip, viewport=5 9 510 340]{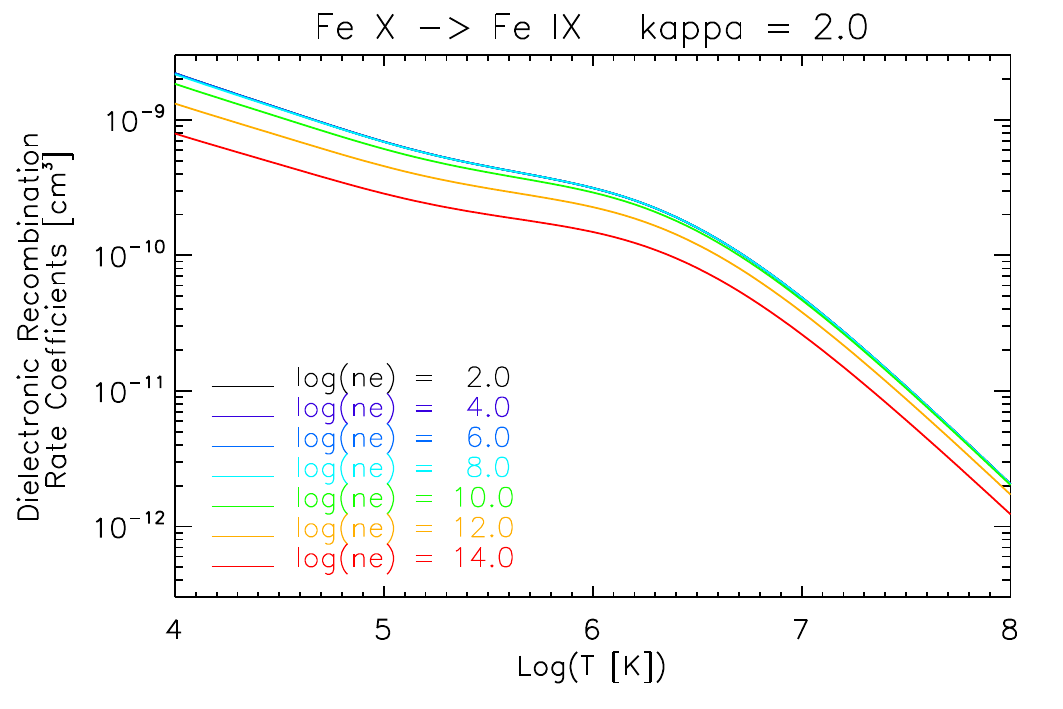}
\caption{Dielectronic recombination rate coefficients where the effects of finite density on dielectronic recombination is shown for Maxwellian distribution (left) and $\kappa$-distribution with $\kappa$\,=\,2 (right). Several ions are shown, \ion{Si}{5}, i.e., recombination from \ion{Si}{5} to \ion{Si}{4}; top) \ion{Fe}{8} (middle), and \ion{Fe}{10} (bottom).}
\label{Fig:dr_suppressed}
\end{figure*}
%%--------------------------
%
%%-------------------------- Extra FIGURE 6  -NEW
\begin{figure*}[t]
  \centering     
    \includegraphics[width=8.8cm, clip, viewport=5 9 545 340]{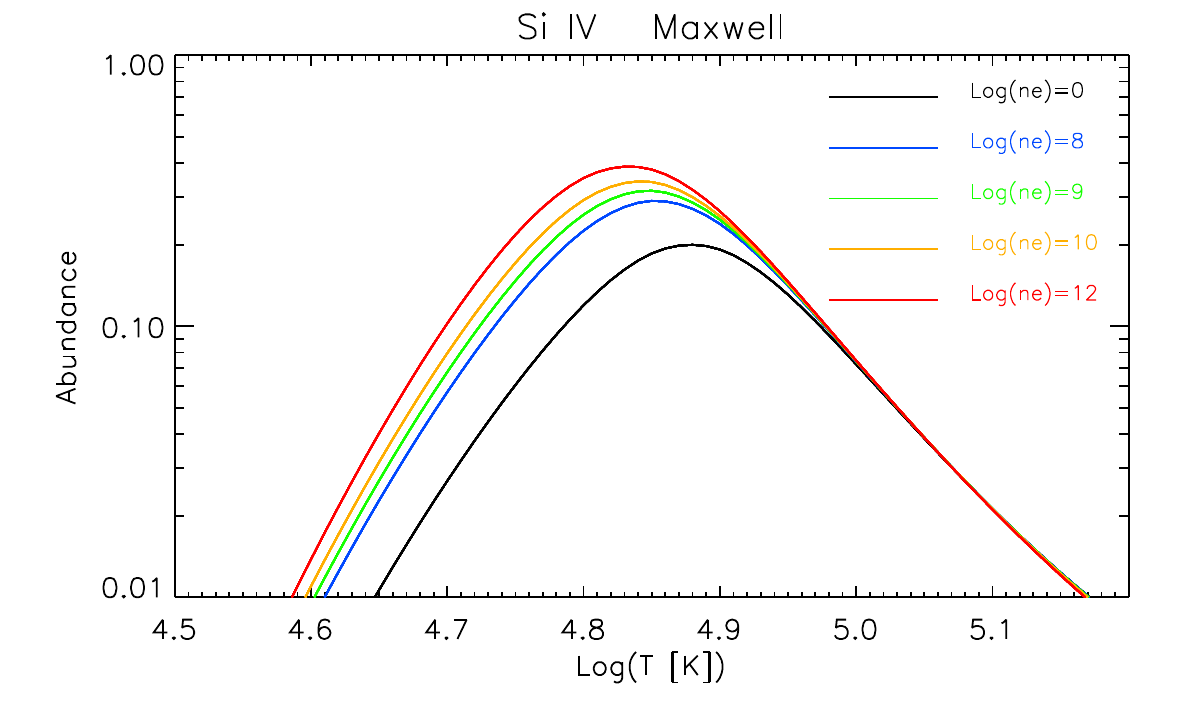}
    \includegraphics[width=8.8cm, clip, viewport=5 9 545 340]{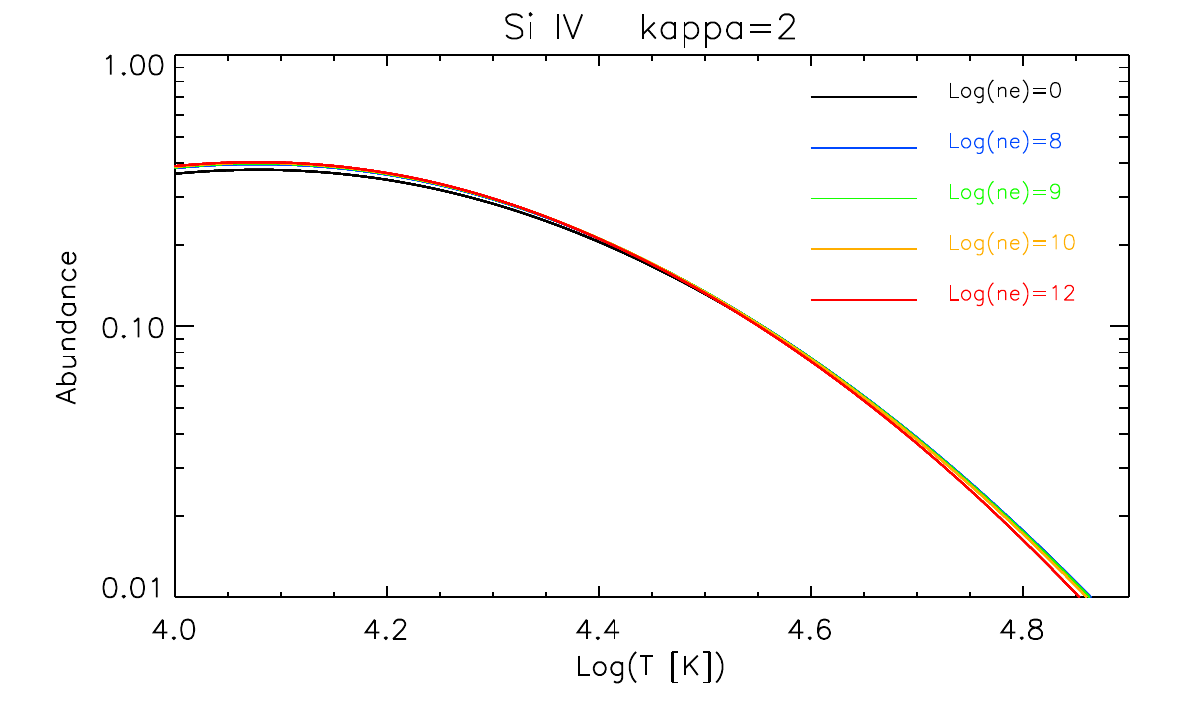}
    \includegraphics[width=8.8cm, clip, viewport=5 9 545 340]{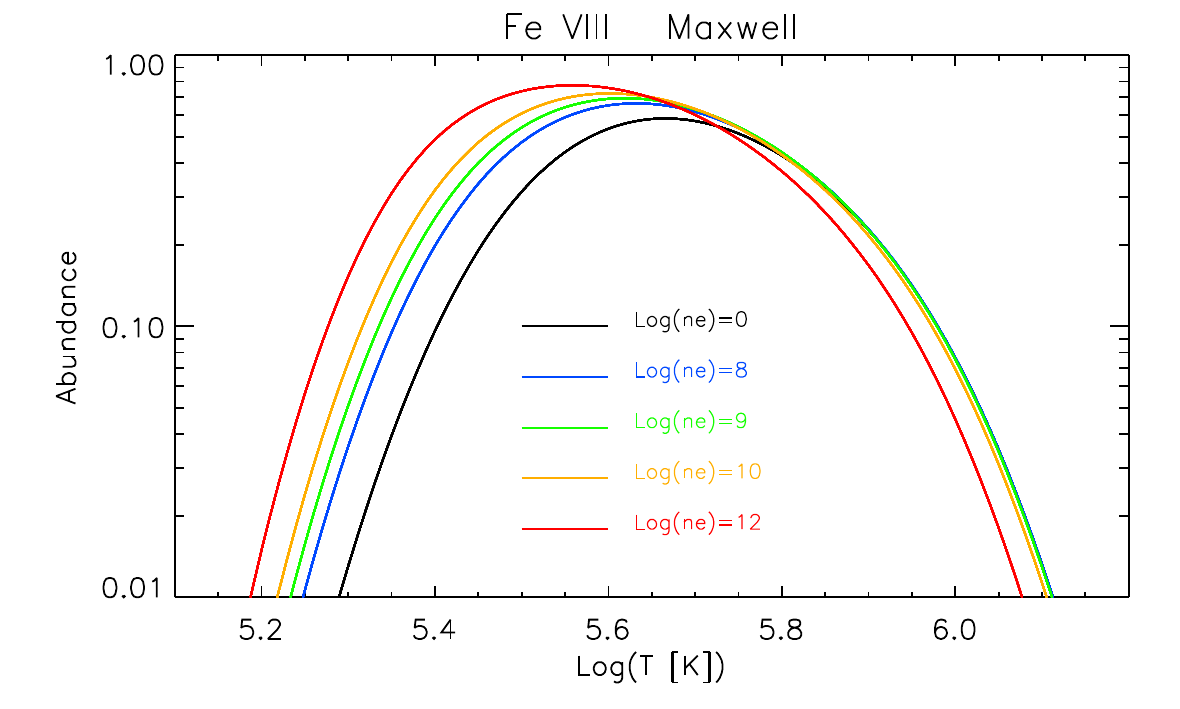}
    \includegraphics[width=8.8cm, clip, viewport=5 9 545 340]{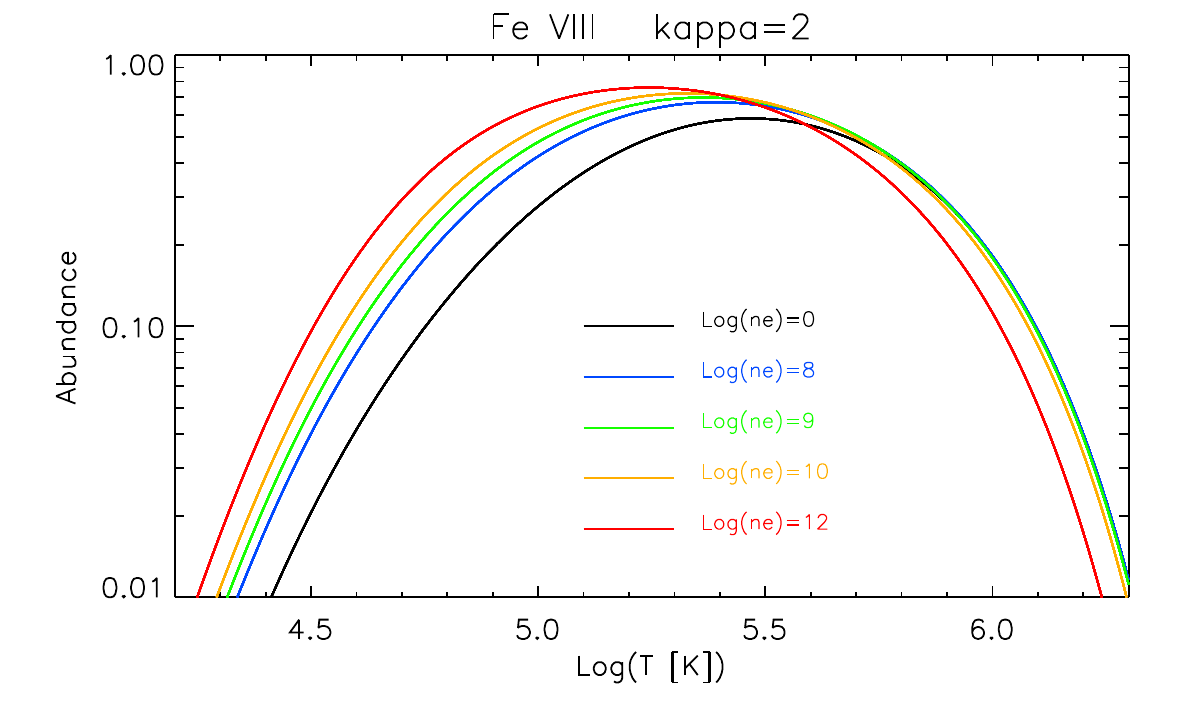}
    \includegraphics[width=8.8cm, clip, viewport=5 9 545 340]{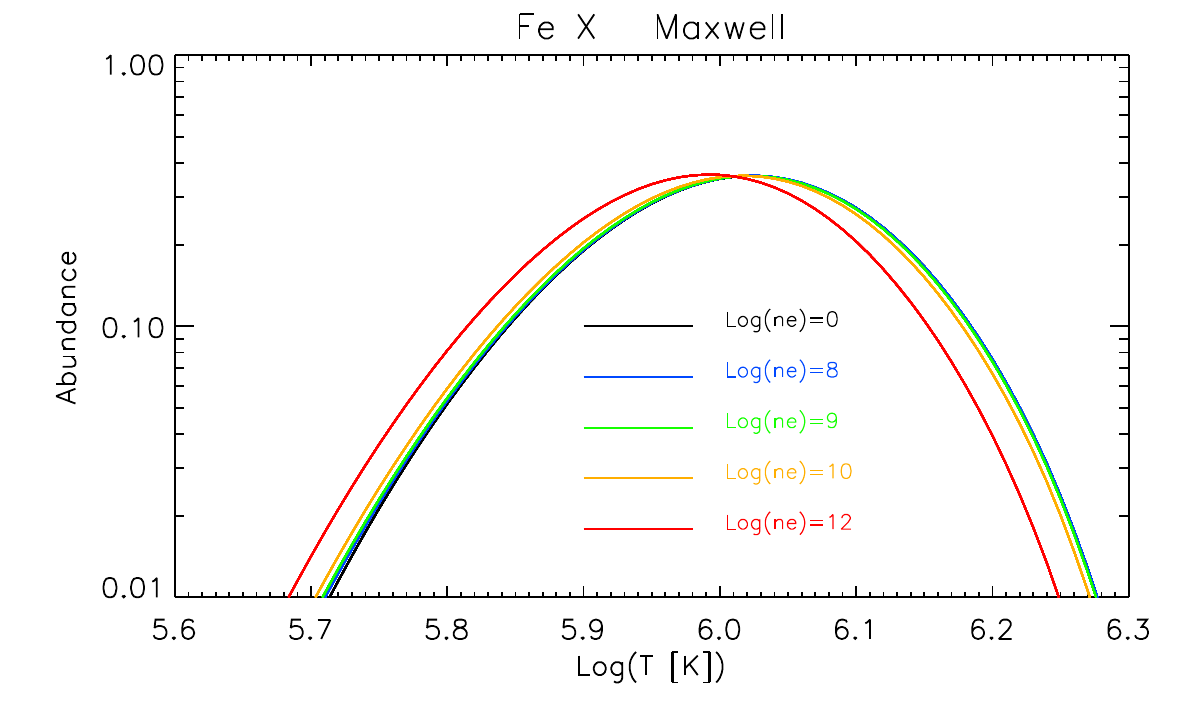}
    \includegraphics[width=8.8cm, clip, viewport=5 9 545 340]{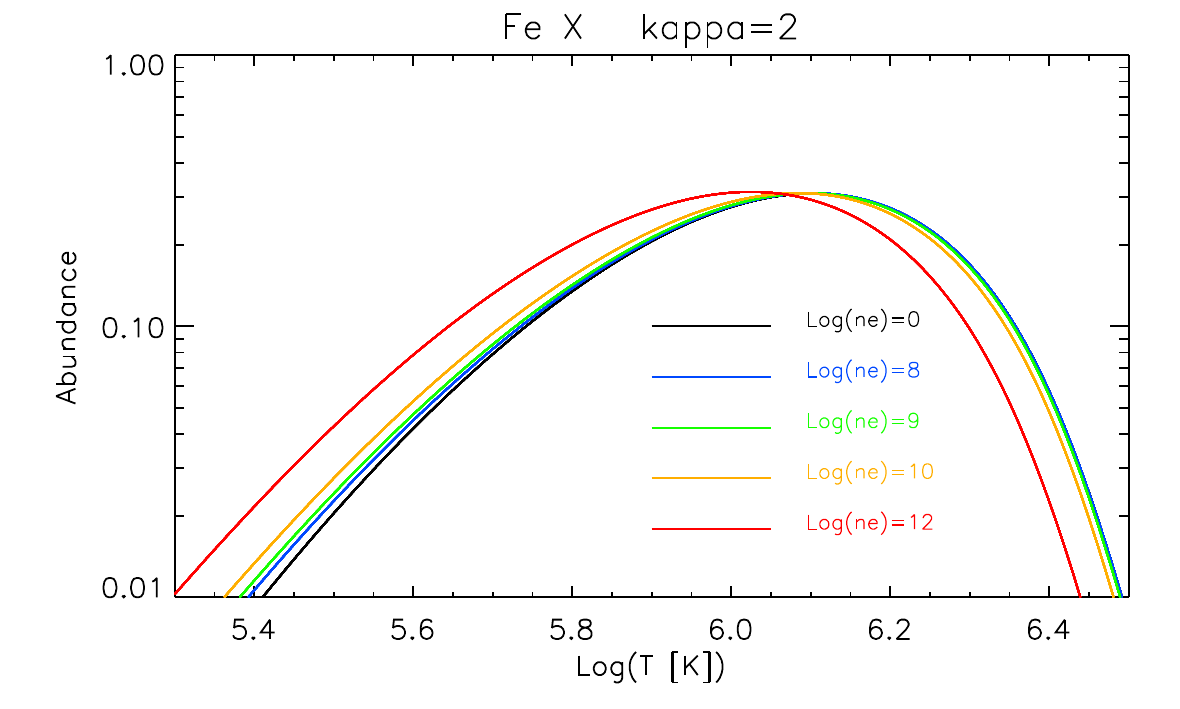}
\caption{Relative ion abundances of \ion{Si}{4} (top), \ion{Fe}{8} (middle), and \ion{Fe}{10} (bottom) where the effects of finite density on dielectronic recombination are shown for Maxwellian distribution (left) and $\kappa$-distribution with $\kappa$\,=\,2 (right). Calculations for all distributions, including the Maxwellian, were based on the ionization cross-sections of \citet{Hahn17} and recombination rates from CHIANTI version 10.1.}
\label{Fig:single_ion}
\end{figure*}
%%--------------------------
%
%%-------------------------- FIGURE 7 (was Figure 5 in submitted version)
\begin{figure}[t]
  \centering     
    \includegraphics[width=17.5cm, clip, viewport=10 30 885 260]{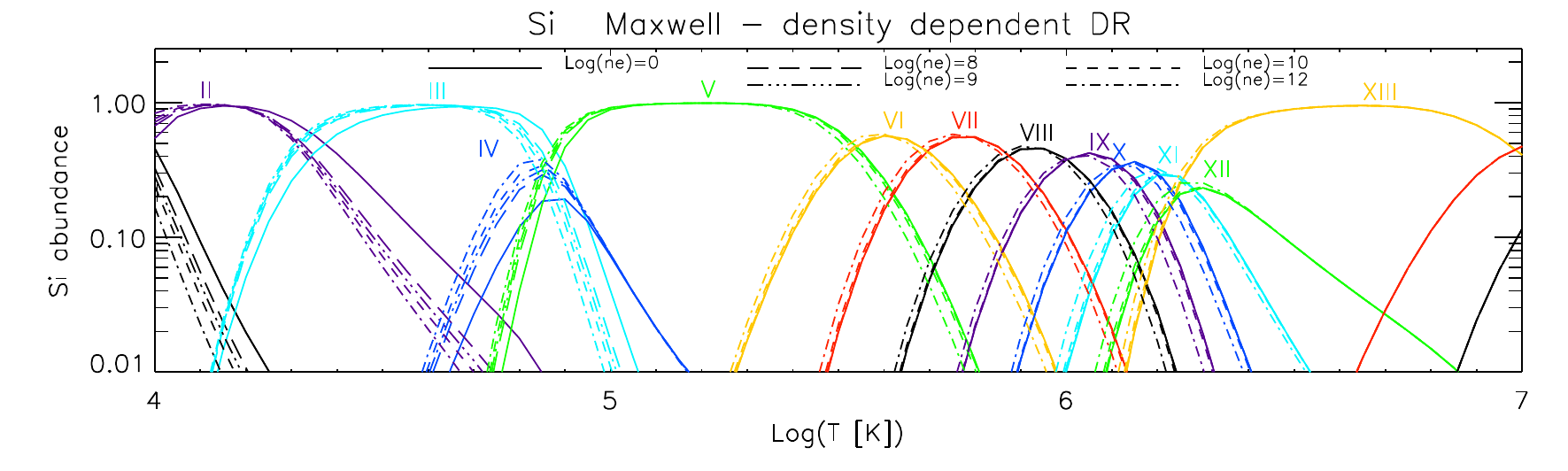}
    \includegraphics[width=17.5cm, clip, viewport=10 30 885 260]{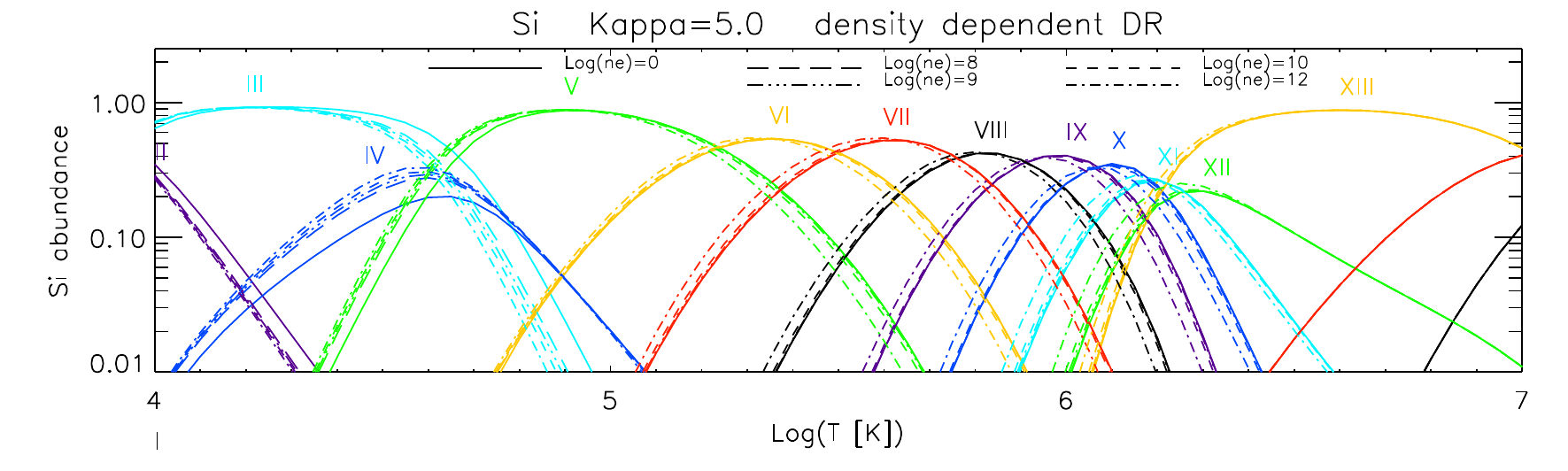}
    \includegraphics[width=17.5cm, clip, viewport=10 30 885 260]{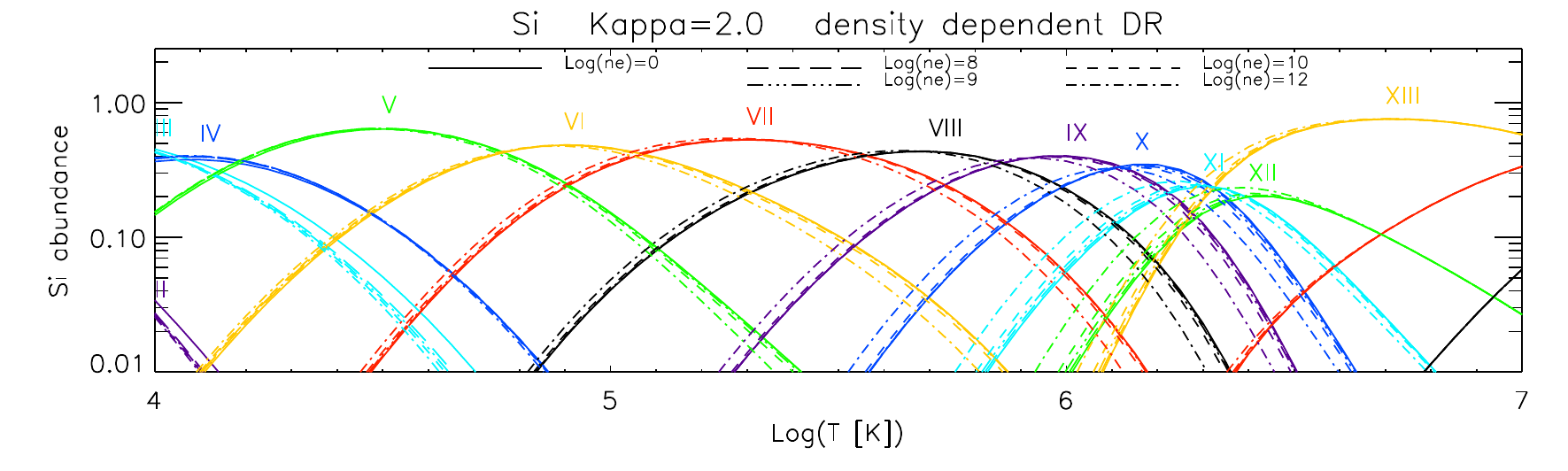}
    \includegraphics[width=17.5cm, clip, viewport=10 12 885 260]{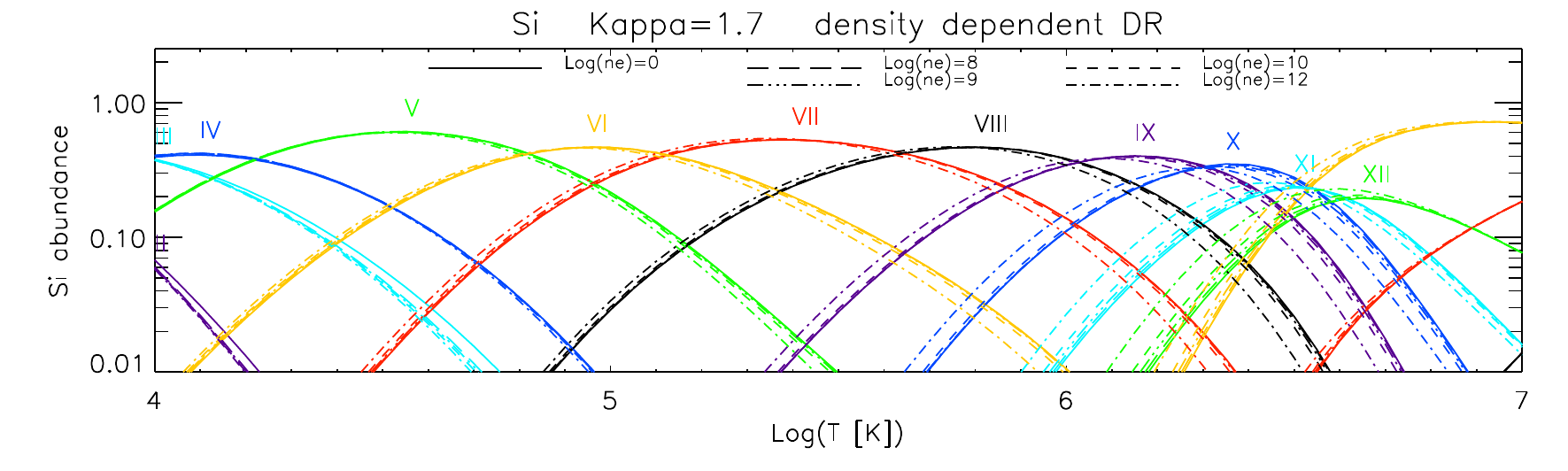}
    \caption{Silicon relative ion abundances (ionization equilibrium) for Maxwellian distribution (top) and $\kappa$-distribution with $\kappa$=5, 2, and 1.7 (rows 2--4). Individual linestyles correspond to different values of \logne. Calculations for all distributions, including the Maxwellian, were based on the ionization cross-sections of \citet{Hahn17} and recombination rates from CHIANTI version 10.1. Note that the effect of EIMI was included.}
    \label{Fig:Si_multi_dens}
\end{figure}
%%--------------------------
%
%%-------------------------- FIGURE 8 (was Figure 6 in submitted version)
\begin{figure}[t]
  \centering     
    \includegraphics[width=17.5cm, clip, viewport=10 30 885 260]{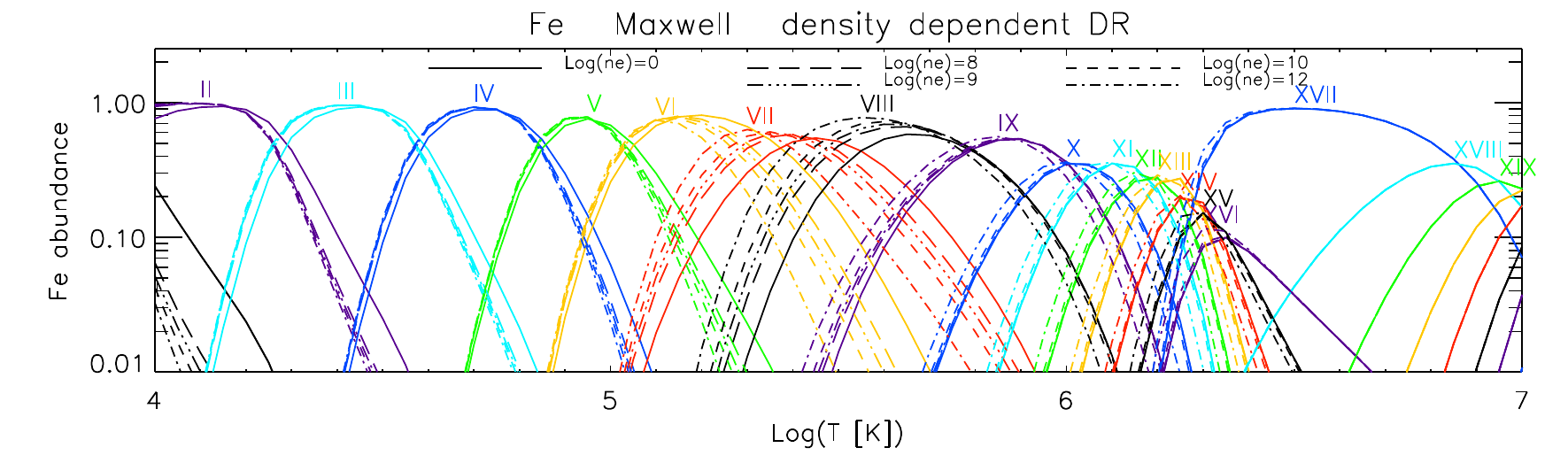}
    \includegraphics[width=17.5cm, clip, viewport=10 30 885 260]{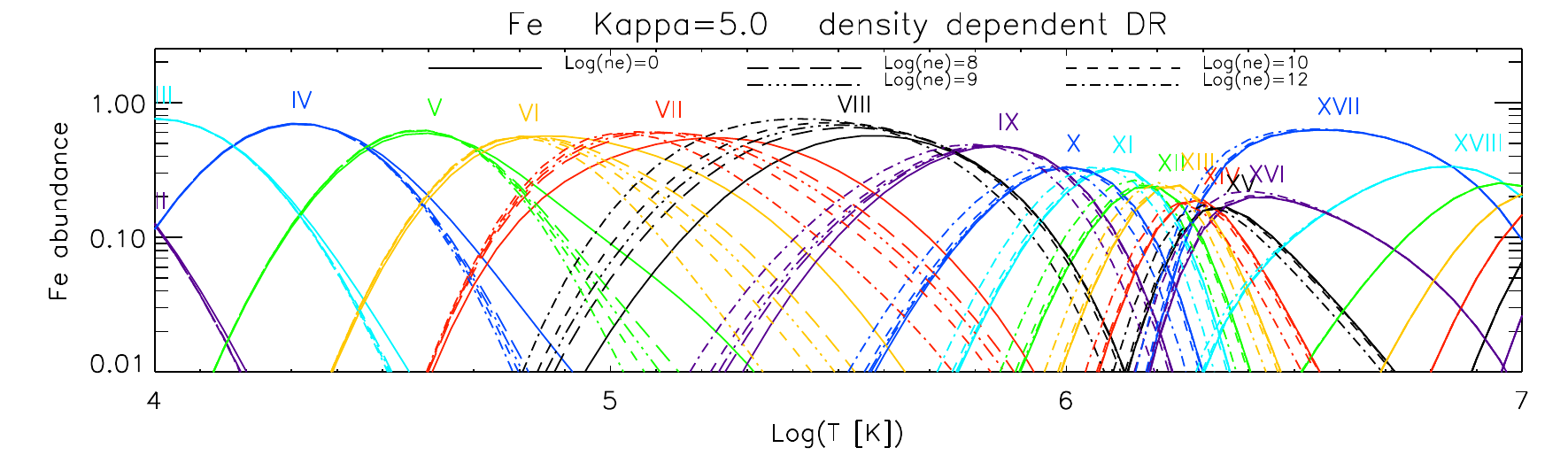}
    \includegraphics[width=17.5cm, clip, viewport=10 30 885 260]{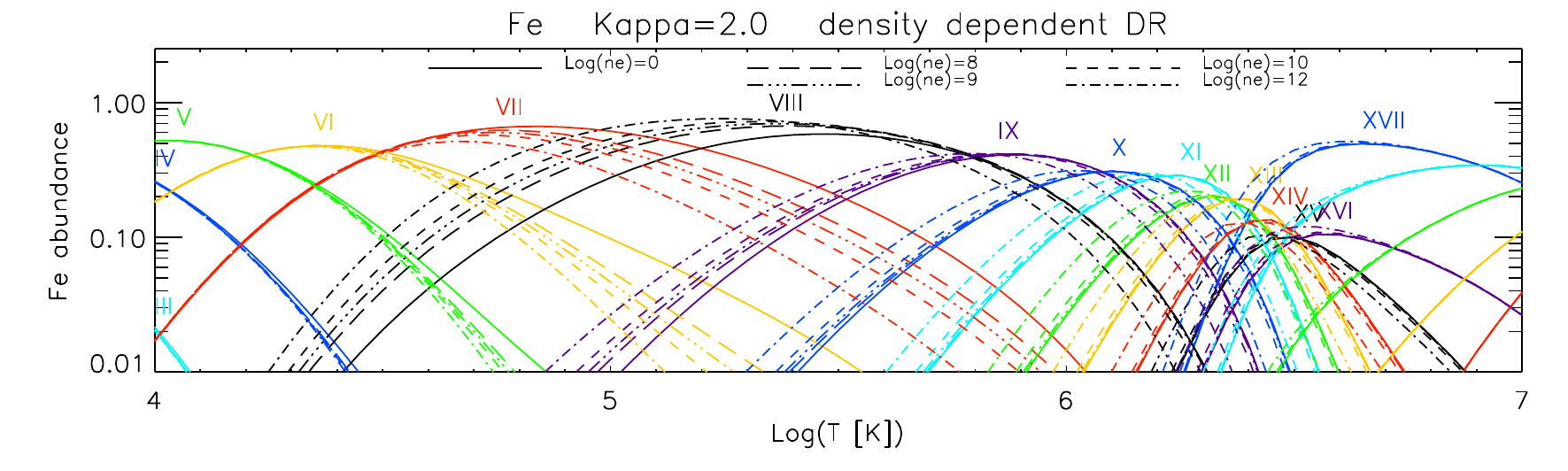}
    \includegraphics[width=17.5cm, clip, viewport=10 12 885 260]{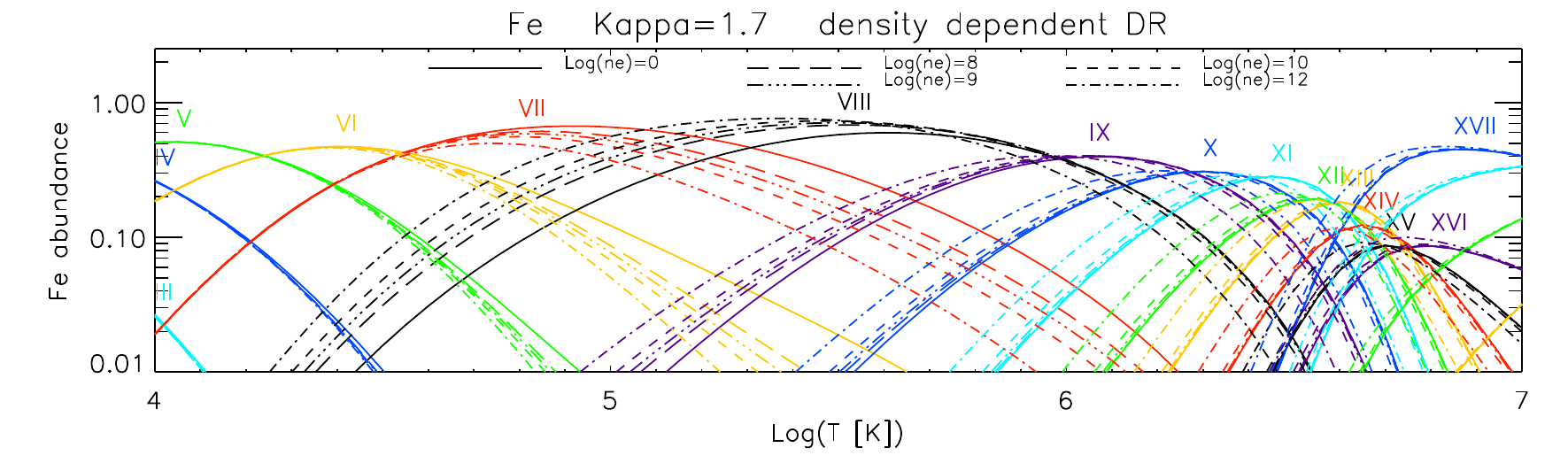}   
    \caption{Same as in \ref{Fig:Si_multi_dens}, but for iron.}
    \label{Fig:fe_multi_dens}
\end{figure}
%--------------------------
%
%-------------------------- FIGURE 9 (was Figure 7 in submitted version)
\begin{figure*}[t]
  \centering     
    \includegraphics[width=17.5cm, clip, viewport=10 12 855 300]{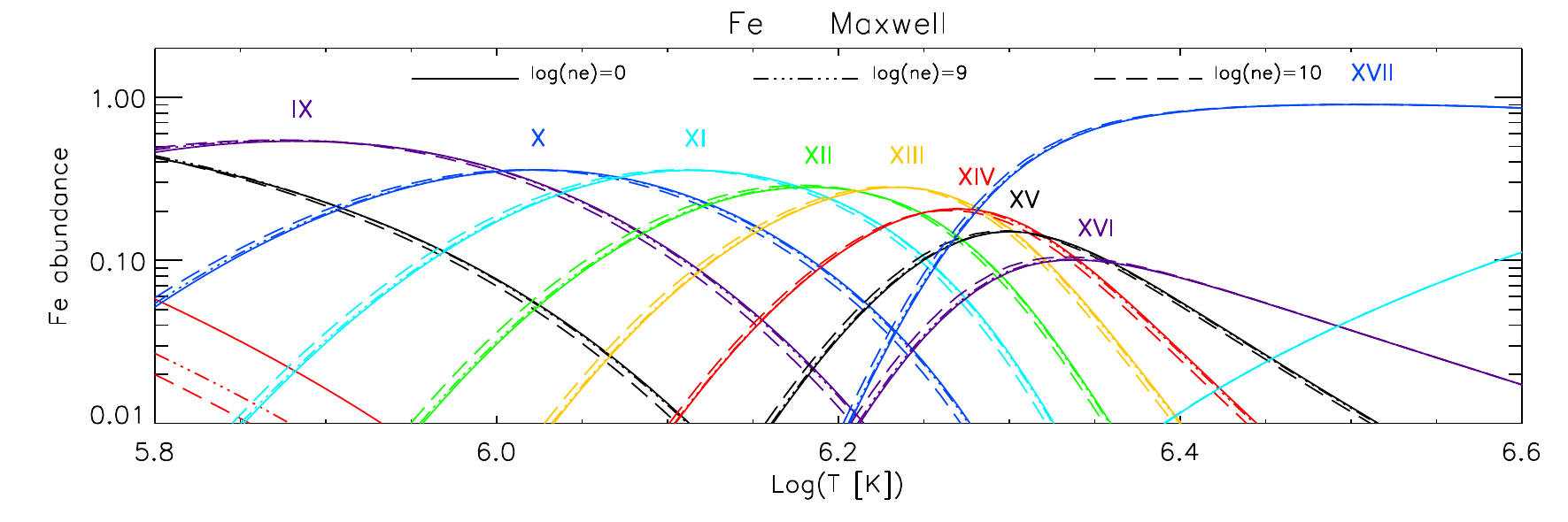}
    \includegraphics[width=17.5cm, clip, viewport=10 12 855 300]{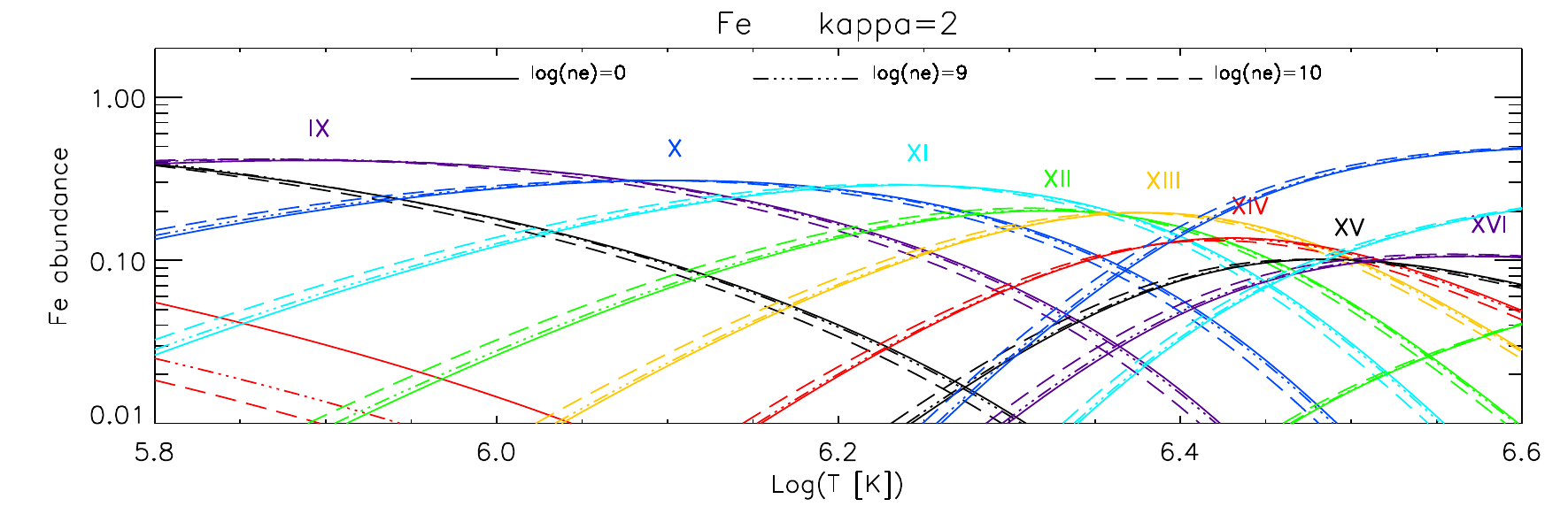}
  \caption{ Fe Maxwellian ionization equilibrium (top) and $\kappa$\,=\,2 (bottom) in the vicinity of T=10$^6$ K for the low electron density (full lines) ans density dependent equilibria for \logne\,=\,9 (dot-dot-dot-dashed lines) and 10 (dashed lines). Calculations for all distributions, including the Maxwellian, were based on the ionization cross-sections of \citet{Hahn17} and recombination rates from CHIANTI version 10.1.}
\label{Fig:fe_mxw_k2_9_10_dens}
\end{figure*}
%--------------------------
%
%-------------------------- FIGURE 10 (was Figure 8 in submitted version)
\begin{figure*}[t]
  \centering     
    \includegraphics[width=17.6cm, clip, viewport=20 10 835 340]{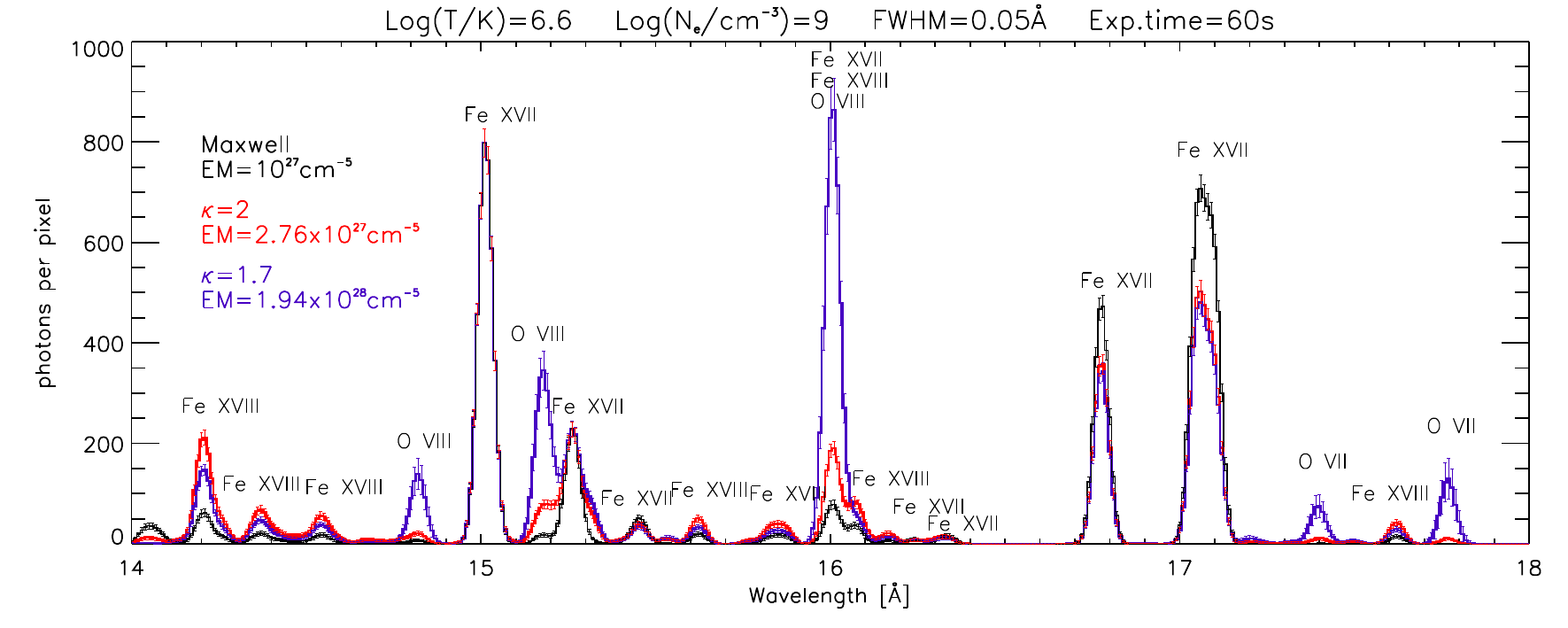}
  \caption{Simulated MaGIXS spectra in the 14--18\,\AA~range showing multiple \ion{Fe}{17}, \ion{Fe}{18}, \ion{O}{7}, and \ion{O}{8} lines at constant \logt\,=\,6.6. Maxwellian spectra are denoted by black lines, while $\kappa$\,=\,2 and 1.7 are denoted by red and violet, respectively. Note the spectra are scaled to keep the \ion{Fe}{17} 15.01\,\,\AA~/\,15.26\,\AA~ratio constant to highlight changes in other spectral lines.}
\label{Fig:spectrum_magixs}
\end{figure*}
%--------------------------
%
%
%-------------------------------
\subsection{Density suppression of dielectronic recombination}
\label{Sect:Ioniz_DR_suppression}

Recombination in the outer solar atmosphere consists of two processes, radiative and dielectronic recombination. \citet{Burgess64} has shown that the latter one can be the more important one. Therefore, accurate implementation of the dielectronic recombination is critical for analysis of the observed transition region and coronal spectra. The KAPPA database contains recombination rates for $\kappa$-distributions calculated using the approximate methods developed by \citet{Dzifcakova92} and summarized in Section of 3.1 of \citet{Dzifcakova13}. These rates are valid, similarly as the corresponding Maxwellian ones in CHIANTI, only in the limit of low electron densities \logne\,$\to$\,0.

However, in relatively high-density plasmas, additional electrons within the plasma can lead to electron-ion collisions and ionization from the doubly excited resonance states. Thus, the radiative rate from the doubly excited states are diminished and the total dielectronic recombination rate $R_\mathrm{DR}$ is suppressed. This suppression of dielectronic recombination was studied already by \citet{Burgess69}, \citet{Summers72,Summers74}, and later by \citet{Summers83}, \citet{Badnell93,Badnell03}, \citet{Nikolic13,Nikolic18}, \citet{Young18}, \citet{Dufresne19}, and \citet{Dufresne20,Dufresne21}. In the last three of these works, the generalized collisional-radiative modelling was employed for carbon \citep{Dufresne19}, oxygen \citep{Dufresne20}, and then for low-charge states generally \citep{Dufresne21}. The generalized radiative-collisional modelling incorporates not only density suppression of dielectronic recombination, but also other processes affecting the ion population, such as photo-induced processes and charge transfer. Although these processes can be of importance for transition region ions, such modelling relies on a vast quantity of reliable atomic data that are not yet readily available. Implementing the generalized radiative-collisional modelling would also mean significant divergence from the methods of calculating synthetic spectra employed in present version 10.1 of the CHIANTI database, on which the KAPPA package is based. Therefore, for the present, we focus on studying the behavior of the density suppression of dielectronic recombination with $\kappa$. Although this process is important, we caution the reader the calculations presented below are not a substitute for generalized collisional radiative modelling. 

Generally, the suppression of dielectronic recombination can be expressed through a dimensionless factor of $S^M(T,N_\mathrm{e},q)$ \citep[][]{Nikolic13,Nikolic18}:
\begin{equation}
   R_\mathrm{DR}(N_\mathrm{e},T,q,M)= R_\mathrm{DR}(T) \times S^M(N_\mathrm{e},T,q),
   \label{Eq:DR_suppression}
\end{equation}
where $R_\mathrm{DR}(T)$ is the dielectronic recombination rate in the \logne\,$\to$\,0 limit, $R_\mathrm{DR}(N_\mathrm{e},T,q,M)$ is the density-suppressed rate, $q$ is a parameter depending on the ion, and $M$ is the isoelectronic sequence. \citet{Nikolic13} used earlier generalized radiative collisional models to develop an approximation  formula for the suppression factor as a function of isoelectronic sequence, charge, electron density, and temperature. \citet{Nikolic18} presented improved fits to calculate suppression of dielectronic recombination at intermediate electron densities, but only for the Maxwellian electron distribution. These authors have shown that the suppression factor depends on the atomic parameters $q$ of ion, on the electron density \dens, and through activation density on $T^{1/2}$ \citep[Equation 3 of][]{Nikolic18}. Therefore, the effect of electron density dominates over the effect of electron temperature. %(and thus also over the effect of electron distribution).  

As no attempt to calculate suppression factors of dielectronic recombination for any other distribution than Maxwellian was made so far, we calculated the suppression of dielectronic recombination $S_{\kappa}^M(N_\mathrm{e},T,q)$ for $\kappa$-distributions. To that, we employed the Maxwellian decomposition approach of \citet[][]{Hahn15a}. These authors approximated $\kappa$-distributions by a sum of several Maxwellians with different temperatures $T_j$,
\begin{equation}
    f_\kappa(E;T)dE = \sum_{j} a_j \, f_\mathrm{Maxw}(E;T_j)dE\,.
    \label{Eq:Maxw_decomp_kappa}
\end{equation}
This allowed them to use the linearity property and thus calculate the non-Maxwellian rates $R_\kappa(T)$ for any collisional process as a weighted sum of the corresponding Maxwellian rates \citep[see Equation 12 of][]{Hahn15a}:
\begin{equation}
    R_\kappa(T) = \sum_{j} a_j \, R_\mathrm{Maxw}(T_j)\,,
    \label{Eq:Maxw_decomp_rates}
\end{equation}
where the coefficient $a_j$ and temperatures $T_j$ are tabulated in \citet{Hahn15a}. It follows that the suppression factor for $\kappa$-distributions can then also be calculated as a weighted sum of Maxwellian suppression factors:   
\begin{equation}
	S_{\kappa}^M(N_\mathrm{e},T,q)=\sum_j a_j S^M_\mathrm{Maxw}(N_\mathrm{e},T_j,q),
	\label{Eq:SF}
\end{equation}
The suppressed dielectronic recombination rate for $\kappa$-distributions is then simply given by the expression analogous to Equation (\ref{Eq:DR_suppression})
\begin{equation}
        R_{\mathrm{DR},\kappa}(N_\mathrm{e})= R_{\mathrm{DR},\kappa} \times S^M_\kappa(N_\mathrm{e},T,q)\,,
        \label{Eq:DR_suppression_kappa}
\end{equation}
where $R_{\mathrm{DR},\kappa}$ is dielectronic recombination rate for $\kappa$-distributions in the low-density limit. We note that although the present method is based on fits of \citet{Nikolic18} and thus not exact, it allows us to estimate the effect of the electron density on the ionization equilibrium for $\kappa$-distribution.

We found that the suppression factors $S_{\kappa}^M(N_\mathrm{e},T,q)$ for $\kappa$-distributions are usually very close to the Maxwellian ones at the same temperature. Differences reach only about 10\% in most cases, although larger differences can occur as discussed below. Generally, the differences are largest for low values of $\kappa$. Examples of the behavior of the suppression factors $S_{\kappa}^M(N_\mathrm{e},T,q)$ with $\kappa$ are shown in Figure \ref{Fig:dr_sf}. The cases shown include dielectronic recombination from \ion{Si}{4} to \ion{Si}{3}, as well as \ion{Fe}{8}\,$\to$\,\ion{Fe}{7}, \ion{Fe}{10}\,$\to$\,\ion{Fe}{9}, and \ion{Fe}{15}\,$\to$\,\ion{Fe}{14}. Where the Maxwellian suppression factor increases with temperature, the changes with $\kappa$ are small, and the suppression factor for low $\kappa$ is slightly smaller than for Maxwellian distribution. This behavior with $\kappa$ is independent of electron density. For ions such as \ion{Fe}{8} and \ion{Fe}{10}, this behavior of the suppression factor occurs in the entire range of $T$ (see Figure \ref{Fig:dr_sf}), while for ions such as \ion{Fe}{15}, this behavior occurs at temperatures \logt\,$\gtrapprox$\,6, where the ion is formed. Clearly, the suppression factor in these cases depends primarily on \logne~and only weakly on $\kappa$. These small differences in the suppression factors with $\kappa$ are however comparable with precision of the fits to the suppression factors themselves as obtained by \citet{Nikolic18}. 

Larger differences with $\kappa$ in the suppression factors occur in cases where the Maxwellian suppression factors decrease with $T$. With decreasing $\kappa$, the suppression factor progressively increases, i.e., the dielectronic recombination rate becomes progressively less suppressed (cf., Equation \ref{Eq:DR_suppression_kappa}). This behavior of the suppression factor is important for transition region ions such as \ion{Si}{4} at temperatures where the ion is formed and occurs for a range of electron densities (see the top panel of Figure \ref{Fig:dr_sf}). It also occurs at high densities for ions such as \ion{Fe}{15}, but only at relatively low temperatures below \logt\,$\lessapprox$\,6, where the ion is not formed in ionization equilibrium.

Figure \ref{Fig:dr_suppressed} shows the resulting density suppressed dielectronic recombination rates for Maxwellian (left panels) and $\kappa$-distribution with $\kappa$\,=\,2 (right panels). Other $\kappa$-distributions are not shown, as in most cases the suppression coefficient is not strongly sensitive to $\kappa$. It can be seen that the density suppression strongly depends on \logne, the atomic parameters of ions (with \ion{Fe}{8} being much more strongly affected than \ion{Fe}{10}), and only slightly on $T$. 

How the resulting relative ion abundances are affected varies depending on all parameters, mostly on the ion and \dens, but also $T$ and $\kappa$, as the changes in the low-density ionization equilibrium with $T$ and $\kappa$ \citep[see][]{Dzifcakova13} are compounded or reduced by changes with \dens~for the individual ion. Generally, the peaks of the relative ion abundances are shifted to lower \logt~for higher electron densities. Elements with higher $Z$ are typically more affected, and largest changes can occur at transition region temperatures, i.e., for \logt\,$\lesssim$\,6.

A comparison of the resulting ionization equilibria (i.e., relative ion abundances) for Si and Fe are shown in Figures \ref{Fig:single_ion}, \ref{Fig:Si_multi_dens}, and \ref{Fig:fe_multi_dens}, respectively. Figure \ref{Fig:single_ion} shows the behavior of the relative ion abundance for several important ions, \ion{Si}{4}, \ion{Fe}{8}, and \ion{Fe}{10}, while the Figures \ref{Fig:Si_multi_dens} and \ref{Fig:fe_multi_dens} show multiple ions of Si and Fe, respectively. Aside from the density-dependent DR, these ionization equilibria also contain the effects of EIMI (Section \ref{Sect:Ioniz_EIMI}). As already noted however, other processes could still influence the ionization equilibrium, especially for low charge states \citep[see for example][]{Dufresne20}. 

For the elements Si and Fe, the resulting ionization equilibrium in the transition region is sensitive to the electron density regardless of the electron distribution, with some ions such as \ion{Si}{4} and \ion{Fe}{6}--\ion{Fe}{8} being affected more than others. For example, at electron densities \logne\,$\approx$\,10--11 typical of the transition region, the peak of \ion{Fe}{7} is shifted to lower $T$ by a factor of 0.7 both for the Maxwellian and $\kappa$\,=\,2, and the peaks get wider with progressively lower $\kappa$ (a well-known effect, see Figure \ref{Fig:fe_multi_dens} and \citet{Dzifcakova13}.) The relative abundance of \ion{Si}{4} is very sensitive to \dens, but more so for Maxwellian than for $\kappa$\,=\,2. For the Maxwellian distribution, the peak of \ion{Si}{4} is shifted slightly to about \logt\,=\,4.8, and at densities of \logne\,=\,10 it is nearly two times higher than for the low-density limit \citep[see also Figure 13 of][]{Polito16}. For $\kappa$\,=\,2 however, the peak of \ion{Si}{4} is almost independent of \dens. This is due to the increase in the suppression factor for such low $\kappa$ (see top panel of Figure \ref{Fig:dr_sf}).

At temperatures and electron densities typical of the solar corona, the effects of suppression of dielectronic recombination at higher \dens~are much more subtle. Regardless of the value of $\kappa$, for \logne\,=\,9--10 the shift of the ionization peaks with \dens~is small and the peaks almost do not change their shapes (see Figure \ref{Fig:fe_mxw_k2_9_10_dens}). This is important for diagnostic purposes, as the previous diagnostics of $\kappa$ \citep{Dudik15,Lorincik20,DelZanna22} were based on the low-density ionization equilibrium calculations. For flare temperatures and densities, the density suppression of dielectronic recombination is negligible and thus does not affect the diagnostics of flaring plasma \citep[cf.,][]{Dzifcakova18}.

We note that the CHIANTI software and database as of yet does not contain ionization equilibria with density suppressed dielectronic recombination or EIMI. Therefore, the respective Maxwellian ionization equilibria are also included in the present version of the KAPPA database. The naming conventions for the respective ionization equilibrium filenames are detailed in Appendix A. Finally, we caution the reader that the effects of finite density due to suppression of dielectronic recombination and the effects of $\kappa$-distributions may not be readily distinguishable, especially in the transition region, and that caution should be exercised in interpreting the observed spectra arising from plasma at electron densities where these effects play a role.

%
%_________________________________________
\section{Excitation Rates}
\label{Sect:Excit_rates}

We endeavor to maintain the database compatible with the latest version of CHIANTI, currently in version 10.1\footnote{Database available directly at \url{https://db.chiantidatabase.org/}}. Following that, there are no major changes in the excitation data within KAPPA since the Paper II. 
The only exception is that the KAPPA database now includes the requisite containing collisional excitation and deexcitation rates for the respective values of $\kappa$\,=1.7, 1.8, 1.9, as well as 2.5 (see also Section \ref{Sect:Ioniz_low_kappa}) so that the respective synthetic spectra can be calculated. The naming conventions for the corresponding files are described in Appendix B.

We note that the atomic datasets are huge, with some ions containing hundreds of energy levels. Therefore, maintaining compatibility with CHIANTI is a huge task. As the atomic data within CHIANTI can change at any time, KAPPA contains since Paper II its own branch of Maxwellian excitation cross-sections for spectral synthesis. This is done so that the corresponding Maxwellian calculations are always available. Should the compatibility of the NMED data within KAPPA with respect the atomic data in CHIANTI be broken at any time, the users are encouraged to contact the KAPPA team with a request to update our atomic data for the $\kappa$-distributions.

As an example of the synthetic spectrum calculated for low values of $\kappa$ for multiple ions, in Figure \ref{Fig:spectrum_magixs} we show a portion of the X-ray spectrum at 14--18\,\AA~observable by the MaGIXS instrument \citep[][]{Savage23}, currently scheduled for second launch in 2024. There, three spectra are shown, a Maxwellian one in black, a $\kappa$\,=\,2 one in red, and $\kappa$\,=\,1.7 spectrum in violet color. The spectra are calculated for a constant \logt\,=\,6.6 and scaled in emission measure so that the ratio of two \ion{Fe}{17} lines, 15.01\,\AA~/\,15.26\,\AA~is kept constant. This approach follows the example spectrum provided in Figure 3 of \citet{Dudik19}, and enables one to immediately recognize which spectral lines are sensitive to $\kappa$ if the value of \logt is held constant. Figure \ref{Fig:spectrum_magixs} shows that at this temperature typical of active region cores, the \ion{O}{7} and \ion{O}{8} lines become strongly enhanced with respect to the neighboring \ion{Fe}{17} ones, especially for extremely low $\kappa$\,=\,1.7. However, we note that the present spectra are calculated for a simple case of constant \logt\,=\,6.6; a value where the \ion{Fe}{17} abundance is relatively low for $\kappa$\,=\,1.7 (see Figure \ref{Fig:Fe_low_k}). The temperature is typically a parameter determined from observations using a range of synthetic spectrum calculations where the \logt~can vary \citep[see, for example, Figures 14 and 16 of][and the discussion therein]{Savage23}. Nevertheless, our example calculations demonstrate that at least in principle it could be possible to distinguish the extreme case of $\kappa$\,=\,1.7 even from the $\kappa$\,=\,2 using the optically thin spectra of the solar corona.

%
%
%_________________________________________
\section{Summary}
\label{Sect:Summary}

We performed an update of the ionization equilibrium calculations in the KAPPA database together with adding several improvements. These include:
\begin{enumerate}
    \item Extension of the calculations towards low values of $\kappa$\,$<$\,2 and adding $\kappa$\,=\,2.5,
    \item Addition of electron impact multi-ionization (EIMI),
    \item Addition of density suppression of dielectronic recombination.
\end{enumerate}

The extension of the KAPPA database to extremely low values of $\kappa$\,$<$\,2 was also done for the excitation rates, so that full calculations for such values of $\kappa$ can now be performed. This extension of the database was prompted by the recent results indicating that the value of $\kappa$ in the solar transition region, corona, and flares can be quite low \citep{Dudik17,Dzifcakova18,Lorincik20,DelZanna22}.

The process of EIMI is generally not important for the Maxwellian electron distribution, but becomes important for strongly NMED such as those with low values of $\kappa$. Therefore, its inclusion is necessary for proper diagnostics of the electron distribution using emission line intensities formed in neighboring ionization stages \citep[cf.,][]{Lorincik20,DelZanna22}. In agreement with \citet{Hahn15a}, we find that the ionization balance for Fe is significantly modified at \logt\,$\approx$\,6--7, where multiple ions are affected if the value of $\kappa$ is low.

To evaluate the density suppression of dielectronic recombination for $\kappa$-distributions, we followed the approach of \citet{Nikolic18} by calculating the respective suppression factors for $\kappa$-distributions. To do that, we employed the Maxwellian decomposition method of \citet{Hahn15a}. We find that the suppression factors are in most cases very close to Maxwellian. At the same temperature, they are within 10\% of the respective Maxwellian ones, although larger differences can occur especially for transition-region ions such as \ion{Si}{4}. The ionization equilibria were subsequently calculated for all values of $\kappa$ and a range of electron densities, and made available within the KAPPA database. Notably, the suppression of dielectronic recombination affects the coronal iron ions, as well as several transition-region ions, such as \ion{Si}{4}. For the latter, the density suppression of dielectronic recombination becomes relatively unimportant for extremely low values of $\kappa$.

We note that the present improvements are not a substitute for the generalized collisional-radiative modelling that is required for some emission lines such as those from the transition region \citep[see, e.g.,][and references therein, for a detailed discussion]{Dufresne20}. Nevertheless, the processes listed above should serve to increase the accuracy of diagnostics of $\kappa$-distributions, especially low values of $\kappa$, in the solar corona and possibly in other astrophysical environments.

\begin{acknowledgements}
The authors acknowledge support from the Czech Science Foundation, grant No. GACR 22-07155S, as well as institutional support RWO:67985815 from the Czech Academy of Sciences. CHIANTI is a collaborative project involving George Mason University, the University of Michigan (USA), University of Cambridge (UK) and NASA Goddard Space Flight Center (USA). 
\end{acknowledgements}

%-----------------------------
\bibliographystyle{aasjournal}
\bibliography{2023_KAPPA}

\appendix
%\section{Appendix}
%\label{Sect:Appendix}

\section{Conventions for naming of ionization equilibrium files}
\label{Sect:A}

We detail here the naming convention for the ionization equilibrium files. %, for both low values of $\kappa$\,$<$\,2, as well as for the ionization equilibrium files containing density suppressed dielectronic recombination rates.
All filenames begin with the string \texttt{Dz23\_} denoting that the calculations were performed for this paper. The value of $\kappa$ is listed at the end of the filename, followed by the \texttt{ioneq} filename extension. For example, calculations for $\kappa$\,=\,2 are denoted as \texttt{\_kappa\_2.ioneq}.

The ionization equilibrium datasets for decimal values of $\kappa$, both low $\kappa < 2$ as well as $\kappa$\,=\,2.5 are denoted in the filenames as \texttt{\_kappa\_1p7.ioneq} and \texttt{\_kappa\_2p5.ioneq} for $\kappa=1.7$ and 2.5, respectively, and similarly for other non-integer values of $\kappa$. The character \texttt{p} is used to avoid the multiple decimal point characters in the filename.

There are multiple ionization equilibrium files for a single value of $\kappa$, since the density suppression of dielectronic recombination is taken into account. These files were calculated for integer values of \logne\,=\,7--12. These files contain string \texttt{logn} followed by numerical value of the \logne. For example, \texttt{logn10} means \logne\,=\,10. For the ionization equilibria in low-density limit, the corresponding string is \texttt{logn0}. Finally, low-density calculations for $\kappa$\,=\,2 are denoted as \texttt{Dz23\_logn0\_kappa\_2.ioneq}.

\section{Conventions for files containing excitation and deexcitation rates}
\label{Sect:B}

To make the KAPPA database fully able to calculate the synthetic spectra for low $\kappa$\,$<$\,2, the database now contains the collisional excitation and de-excitation rates in the form of \texttt{.ups} and \texttt{.dwns} files for each ion and the respective values of $\kappa$\,=1.7, 1.8, 1.9, as well as 2.5; that is, the values of $\kappa$ that were added to the database (see Section \ref{Sect:Ioniz_low_kappa}). These rates are based on the same approximate atomic cross-sections as for other $\kappa$ values, while the filenames follow the same suffix convention for decimal values of $\kappa$ as described in Appendix A. For example, the file \texttt{fe\_10\_k1p7.ups} contains the excitation rates for \ion{Fe}{10} and $\kappa$\,=\,1.7.

We note that for the $\kappa$-distributions the distribution-averaged collision strengths $\Upsilon_{ij}(T,\kappa)$ for excitation and \rotatebox[origin=c]{180}{$\Upsilon$}$_{ji}(T,\kappa)$ for de-excitation \citep[non-dimensionalized excitation and de-excitation cross-sections, see][and references therein]{Dudik14AA} are not equal, i.e., $\Upsilon_{ij}(T,\kappa)$\,$\ne$\,\rotatebox[origin=c]{180}{$\Upsilon$}$_{ji}(T,\kappa)$. For this reason, there are two separate \texttt{.ups} and \texttt{.dwns} files for each ion. Similarly as in the previous versions of the KAPPA database, the $\Upsilon_{ij}(T,\kappa)$ are scaled, but the \rotatebox[origin=c]{180}{$\Upsilon$}$_{ji}(T,\kappa)$ are not. The scaling factor used for $\Upsilon_{ij}(T,\kappa)$ is equal to $(1 + E/(\kappa -3/2)k_\mathrm{B}T)^\kappa$, where $E$ is the electron energy and $k_\mathrm{B}$ is the Boltzmann constant \citep[see Section 3.2 of][]{Dzifcakova21}.

\end{document}